\documentclass[10pt,a4paper,twocolumn]{article}

\usepackage[utf8]{inputenc}
\usepackage[T1]{fontenc}
\usepackage{amsmath,amssymb,amsfonts}
\usepackage{graphicx}
\usepackage[dvipsnames]{xcolor}
\usepackage[english]{babel}
\usepackage{geometry}
\usepackage{microtype}
\usepackage{listings}
\usepackage{caption}
\usepackage{subcaption}
\usepackage{booktabs}
\usepackage{longtable}
\usepackage{multirow}
\usepackage{array}
\usepackage{float}
\usepackage{textcomp}
\usepackage{balance}

\usepackage[authoryear,round]{natbib}
\usepackage{authblk}

\usepackage{tcolorbox}

\usepackage{hyperref}
\hypersetup{
    colorlinks=true,
    linkcolor=MidnightBlue,
    citecolor=MidnightBlue,
    urlcolor=MidnightBlue
}

\title{The Architecture of Trust: A Framework for AI-Augmented Real Estate Valuation in the Era of Structured Data}

\author[1,*]{Petteri Teikari}
\author[1,2]{Mike Jarrell}
\author[3]{Maryam Azh}
\author[1]{Harri Pesola}

\affil[1]{Mill Hill Garage}
\affil[2]{JB Real Estate Valuation \& Advisory, LLC}
\affil[3]{Atlas Insights}
\affil[*]{Corresponding author: petteri@millhillgarage.com}

\begin{document}

\maketitle

\begin{abstract}
The mandatory implementation of the Uniform Appraisal Dataset (UAD) 3.6 by 2026 represents a fundamental transformation in residential property valuation methodology, transitioning from narrative-based reporting to structured, machine-readable data formats. This paper provides the first comprehensive academic analysis of this regulatory shift within the context of concurrent advances in artificial intelligence, particularly in computer vision, natural language processing, and autonomous systems. Drawing on literature from real estate economics, information systems, and artificial intelligence, we develop a three-layer theoretical framework for AI-augmented valuation systems that addresses both technical implementation and institutional trust requirements. Our analysis reveals that the convergence of regulatory standardization with technological capabilities creates conditions for fundamental market restructuring, with profound implications for professional practice, market efficiency, and systemic risk.

The paper makes several contributions to the literature. First, we document the institutional failures driving transformation, including significant inter-appraiser variability and systematic biases that undermine valuation reliability. Second, we develop a comprehensive architectural framework spanning physical data acquisition, semantic understanding, and cognitive reasoning that integrates emerging technologies while maintaining professional oversight. Third, we address the critical trust requirements for high-stakes financial applications, including regulatory compliance, algorithmic fairness, and uncertainty quantification. Fourth, we propose novel evaluation methodologies that move beyond generic AI benchmarks to domain-specific protocols appropriate for professional services. Our findings suggest that successful transformation requires not merely technological sophistication but careful attention to human-AI collaboration, creating systems that augment rather than replace professional expertise while addressing historical biases and information asymmetries in real estate markets.
\end{abstract}

\section{Introduction}

The United States residential real estate appraisal industry, valued at \$11.3 billion annually (\citep{ibisworld2025}, in 2023), stands at a critical juncture where regulatory evolution intersects with technological innovation. The Government-Sponsored Enterprises' (GSEs) mandate for Uniform Appraisal Dataset (UAD) 3.6 implementation by 2026 represents more than incremental regulatory updating; it constitutes a fundamental reconceptualization of how property valuations are created, transmitted, and utilized within mortgage markets \citep{mac2024, mae2024}. This transformation from narrative-based reporting to structured, XML-formatted data occurs simultaneously with significant advances in artificial intelligence capabilities, creating unique conditions for market evolution that demand careful academic examination.

The theoretical significance of this convergence extends beyond operational efficiency to fundamental questions about information production in financial markets. Traditional appraisal practice, grounded in the three approaches to value---sales comparison, cost, and income---has long relied on tacit knowledge and professional judgment that resists standardization \citep{pagourtzi2003}. As \cite{mooya2016} argues, conventional valuation theory grounded in neoclassical economics inadequately addresses the complexity of real estate markets, where the assumption of market value as a determinate, knowable quantity fails to account for fundamental uncertainty. The imposition of structured data requirements through UAD 3.6 challenges this paradigm, potentially transforming appraisal from craft-based practice to data-driven science.

The industry simultaneously faces severe capacity constraints that threaten its ability to serve growing housing markets. The appraisal workforce confronts a demographic crisis with the average appraiser age exceeding 50 years, declining new entrants despite growing demand, and geographic misalignment between appraiser availability and housing market activity \citep{reggora2023}. These constraints create bottlenecks in mortgage origination, increasing costs and delays throughout the housing finance ecosystem.

Recent developments in artificial intelligence offer potential solutions to these challenges. The emergence of compound AI systems---architectures that orchestrate multiple specialized models to perform complex, multi-step tasks---aligns particularly well with the multifaceted nature of property valuation \citep{raghavan2025, hassan2024}. Unlike monolithic models that attempt to solve all aspects of a problem simultaneously, compound systems decompose complex tasks into manageable components, each addressed by specialized models. For appraisal, this might involve separate models for image analysis, text processing, market analysis, and report generation, orchestrated by an overarching system that maintains consistency and compliance.

The evolution toward agentic AI systems---those capable of perceiving their environment, pursuing goals through multi-step reasoning, and autonomously utilizing tools and APIs---further enhances the potential for meaningful automation \citep{acharya2025, sapkota2025}. These systems can potentially automate time-consuming tasks such as data gathering from multiple sources, comparable property research across fragmented databases, and preliminary analysis while ensuring that critical professional judgment, final value reconciliation, and legal responsibility remain with licensed appraisers.

However, the application of AI to real estate valuation raises profound questions about trust, accountability, and systemic risk that extend beyond technical capabilities. Unlike consumer-facing applications where errors have limited consequences, property valuations directly impact credit decisions, wealth accumulation, and systemic financial stability. The role of flawed valuations in the 2008 financial crisis, as documented by \cite{wu2015}, underscores the importance of careful system design that prioritizes reliability and fairness alongside efficiency.

This paper contributes to multiple literature streams by providing the first comprehensive academic analysis of how AI technologies intersect with evolving regulatory requirements in real estate valuation. We develop a theoretical framework that integrates perspectives from information economics \citep{akerlof1970, stiglitz2000}, professional service automation \citep{susskindFutureProfessionsHow2015a}, and algorithmic accountability \citep{diakopoulos2015}. Our analysis bridges the gap between technological possibility and institutional reality, examining not just what can be automated but what should be automated given the complex web of professional, regulatory, and social considerations surrounding property valuation.

The remainder of this paper is organized as follows. Section 2 examines the institutional context driving transformation, including detailed analysis of market failures, regulatory evolution, and the specific requirements of UAD 3.6. Section 3 develops our three-layer architectural framework for AI-augmented valuation systems, incorporating detailed technical analysis of emerging technologies. Section 4 addresses trust requirements including regulatory compliance, algorithmic fairness, and comprehensive uncertainty quantification. Section 5 proposes evaluation methodologies appropriate for professional AI systems that move beyond generic benchmarks. Section 6 analyzes implementation dynamics, stakeholder impacts, and market evolution patterns. Section 7 concludes with implications for theory, practice, and future research directions.

\section{Institutional Context and Market Failures}

Understanding the current transformation in real estate appraisal requires examining both the institutional evolution of property valuation and the market failures that create impetus for change. This section synthesizes literature on valuation theory, presents empirical evidence of systematic problems in current practice, analyzes the fragmented data landscape that complicates valuation, and examines how UAD 3.6 represents a fundamental shift in regulatory approach while incorporating detailed technical context of the transformation.

\subsection{Theoretical Foundations and Practical Divergence}

Property valuation theory rests on neoclassical economic principles of market equilibrium and rational actor behavior. The three approaches to value---sales comparison, cost, and income---represent different methodologies for estimating the same underlying construct: market value \citep{institute2020}. In perfectly efficient markets with complete information, these approaches should theoretically converge to similar estimates \citep{lushtRealEstateValuation2001}. The sales comparison approach, grounded in the principle of substitution, assumes rational buyers will not pay more for a property than the cost of acquiring a similar substitute. The cost approach estimates value by calculating the cost to construct an equivalent structure, adding land value, and subtracting depreciation. The income approach, central to commercial valuation, treats properties as investment vehicles, estimating value based on the present value of future income streams.

However, substantial literature documents the divergence between these theoretical ideals and practical realities in property markets. \citep{mooya2016} comprehensive critique of conventional valuation theory argues that the assumption of market value as a determinate, knowable quantity fundamentally misunderstands the nature of real estate markets. Drawing on critical realist philosophy, Mooya contends that property markets are open systems characterized by fundamental uncertainty, where the conditions necessary for equilibrium pricing---perfect information, homogeneous products, rational actors---simply do not exist. This philosophical challenge to the foundations of valuation theory is supported by extensive empirical evidence of significant variation in professional practice.

The most compelling empirical evidence comes from \cite{eriksen2019}, whose analysis of 8,531 properties with repeat appraisals provides unprecedented insight into inter-appraiser variability. Their methodology involved identifying properties that had multiple appraisals performed by different appraisers within short time periods (averaging 82 days), allowing for direct comparison of how different professionals assess identical properties. The results prove deeply troubling for the profession's claims to objectivity. When different appraisers evaluate the same property, reported gross living area---ostensibly an objective measurement requiring only a tape measure and basic arithmetic---shows discrepancies exceeding 100 square feet in 25\% of cases (with 59.4\% showing more than $\pm$1\% difference in square footage). Given that price per square foot serves as a fundamental metric in residential valuation, such measurement variance can translate to valuation differences of \$20,000 to \$50,000 depending on local market conditions.

The variability becomes even more pronounced for subjective assessments. Property condition ratings, coded on the standardized C1-C6 scale where each level theoretically has clear definitions, achieved consistency across appraisers only 32.3\% of the time. Design quality ratings showed even worse agreement at 26.2\%. These findings cannot be explained by simple measurement error or minor differences in professional judgment; they reflect fundamental inconsistencies in how appraisers perceive, interpret, and document property characteristics. Such variability undermines the theoretical foundation of appraisal as an objective estimation process and raises serious questions about the reliability of valuations for lending decisions.

Beyond individual variability, research documents systematic biases that pose even greater concerns for market efficiency and fairness. Multiple studies have identified the phenomenon of ``appraisal inflation'' or ``contract price confirmation bias,'' where valuations in purchase transactions disproportionately land at or slightly above the agreed contract price \citep{alexandrov2023, eriksen2019}. This pattern suggests that appraisers, despite professional requirements for independence, are influenced by knowledge of the agreed sale price, potentially viewing their role as confirming rather than independently assessing value.

Fannie Mae's research provides particularly detailed analysis of this phenomenon. Using advanced statistical techniques to control for property characteristics and market conditions, they found that knowledge of the contract price resulted in an additional 23\% of properties appraising at or above the contract price compared to what unbiased models would predict. This isn't merely a statistical curiosity---further analysis revealed that mortgages on homes appraised exactly at the contract price show somewhat higher default rates, suggesting that confirmation bias introduces real credit risk into the mortgage system. The mechanism appears to be that marginal transactions that should fail independent valuation scrutiny instead receive passing grades when appraisers anchor on contract prices.

\subsection{Information Production and Market Structure}

\begin{figure*}
    \centering
    \includegraphics[width=1\linewidth]{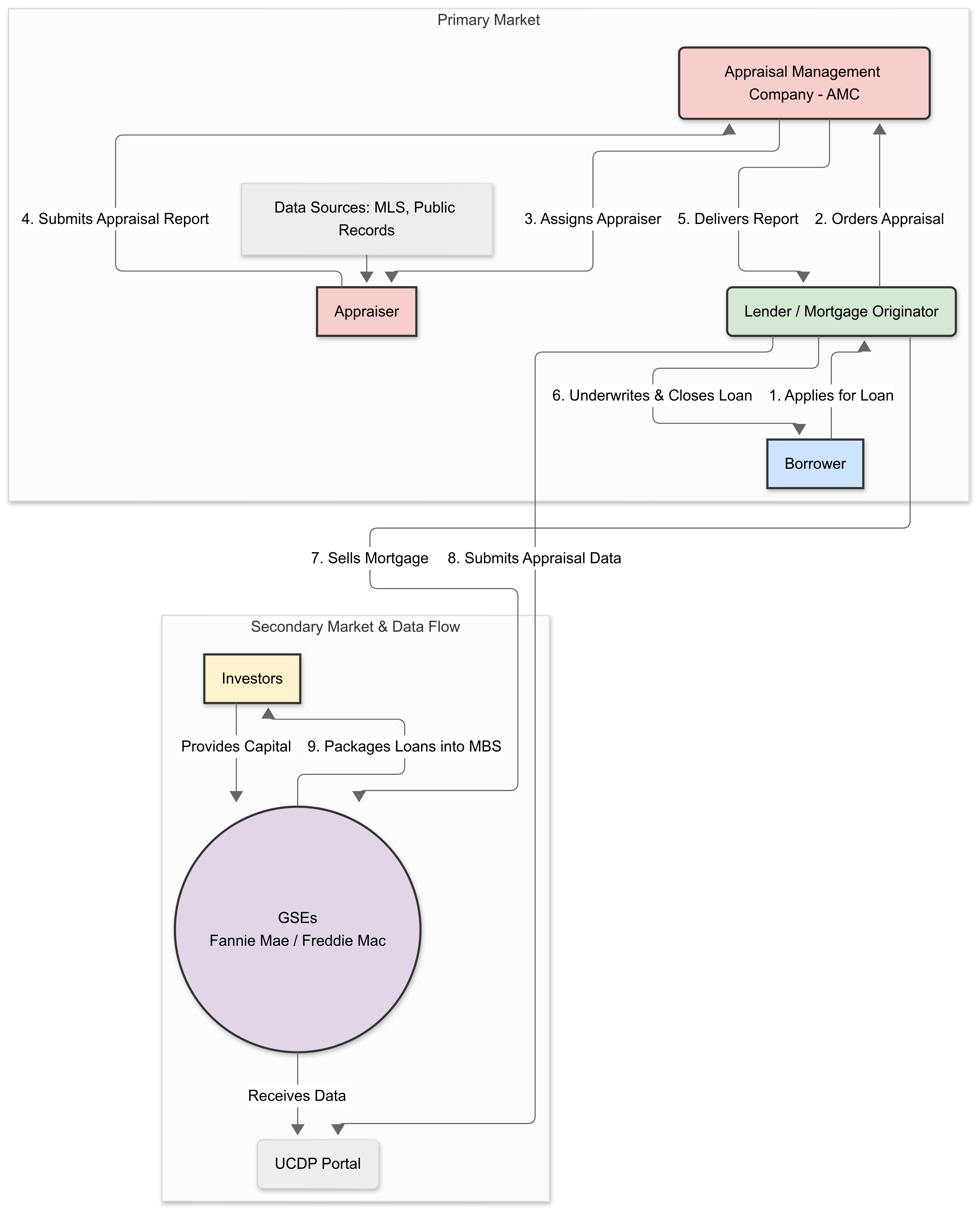}
    \caption{The complex ecosystem of mortgage origination and appraisal, showing the flow from loan application through various valuation methods (traditional appraisal, desktop, hybrid, waivers) to final loan approval, highlighting the different stakeholders and data flows involved}
    \label{fig:mortgage}
\end{figure*}

The fragmented structure of U.S. real estate information systems significantly exacerbates valuation challenges (\autoref{fig:mortgage}), creating what information economists would recognize as a classic market failure in information production. Unlike countries such as the Netherlands or Singapore that maintain centralized, standardized property databases accessible to all market participants, the United States distributes property information across a bewildering array of local and regional systems. More than 500 regional Multiple Listing Services (MLSs) operate as independent entities, each with distinct rules, data formats, access policies, and fee structures \citep{nar2021}. County recording offices---numbering in the thousands---maintain property records, tax assessments, and transaction histories with widely varying levels of digitization and accessibility.

This balkanization creates significant transaction costs for information gathering and introduces multiple sources of error into the valuation process. An appraiser working in a metropolitan area that spans multiple MLS territories must maintain separate subscriptions, learn different search interfaces, and manually compile data from incompatible systems. The search costs alone can consume hours per appraisal, time that could otherwise be spent on analysis and quality control. Moreover, the lack of standardization means that identical property features may be recorded differently across systems---what one MLS calls a ``bonus room'' another might term a ``flex space,'' creating matching and comparison challenges.

From an information economics perspective, this fragmentation represents a classic market failure where the social value of comprehensive, accurate property information far exceeds the private incentives for any individual actor to create such systems \citep{stiglitz2000}. The positive externalities of better information---improved market efficiency, reduced transaction costs, better risk assessment---accrue broadly across market participants, while the costs of creating integrated systems fall on specific actors. This misalignment of costs and benefits helps explain why fragmentation persists despite its obvious inefficiencies.

Commercial data aggregators have partially emerged to address this market failure, with companies like Black Knight (now part of Intercontinental Exchange), CoreLogic, and HouseCanary building businesses around collecting, cleaning, and reselling integrated property data. However, these commercial solutions introduce new forms of information asymmetry \citep{li2023}. Information externalities in real estate markets, data aggregators' proprietary processing methods and selective access policies can create advantages for large institutional players while disadvantaging smaller market participants. The ``black box'' nature of their data cleaning and enhancement processes also raises questions about data quality and potential bias introduction.

The challenges extend beyond simple access to fundamental data quality issues. Industry analysis suggests that approximately 30\% of property listings contain materially inaccurate information \citep{bykowa2022}. Common errors include overstated square footage (often by listing agents seeking to attract buyers), incorrect room counts, undocumented renovations or additions, and misleading property condition descriptions. Each error propagates through the valuation process, as appraisers must either catch these inaccuracies through careful verification or risk basing their valuations on flawed data.

The emergence of automated valuation models (AVMs) since the 1990s represents one market response to these information challenges. The academic literature on AVMs shows a clear evolution from early hedonic pricing models building on \cite{rosen1974}'s theoretical framework, through the application of machine learning techniques in the 2000s, to recent neural network approaches achieving impressive predictive accuracy \citep{kaviani2021}. Modern AVMs using ensemble methods and gradient boosting frameworks like CatBoost can achieve median absolute percentage errors below 5\% in data-rich markets with abundant recent transactions.

However, AVMs' limited adoption for origination purposes reveals persistent concerns about their appropriate use. While they excel at valuing routine properties in active markets, their performance degrades rapidly for unique properties, thin markets, or rapidly changing conditions---precisely where human expertise proves most valuable. More fundamentally, AVMs operate as ``black boxes'' producing valuations without explanations, making them unsuitable for a regulated environment requiring documented reasoning for every conclusion. The inability to explain why an AVM reached a particular value estimate, which adjustments it applied, or how it weighted different factors creates both regulatory compliance challenges and practical difficulties when values must be defended or revised.

\subsection{Regulatory Evolution and the UAD 3.6 Paradigm Shift}

The regulatory framework governing property valuation in the United States reflects ongoing tensions between desires for standardization and consistency versus recognition of professional judgment and local expertise. The Uniform Standards of Professional Appraisal Practice (USPAP), established by the Appraisal Foundation following the savings and loan crisis of the 1980s, attempts to balance these competing demands. USPAP emphasizes the importance of professional judgment while mandating certain procedural requirements, creating what is recognized as a classic professional jurisdiction claim---defining boundaries of expertise while maintaining practitioner autonomy.

The 2008 financial crisis prompted additional regulatory evolution, most notably through the Dodd-Frank Act's appraisal independence requirements. as analyzed by \cite{boehmer2018}, these requirements mandated separation between loan production and appraisal functions, typically through Appraisal Management Companies (AMCs) serving as intermediaries. While intended to reduce pressure on appraisers to hit target values, these requirements added layers of procedural complexity and cost while doing little to address underlying data quality issues or inter-appraiser variability.

\subsection{UAD 3.6: A Fundamental Restructuring}

The Uniform Appraisal Dataset (UAD) 3.6 represents a qualitatively different regulatory approach that addresses information production at its foundation rather than adding procedural overlays. Rather than mandating additional reviews or independence requirements, UAD 3.6 fundamentally restructures how appraisal information is captured, formatted, and transmitted \citep{mae2024a, mac2024}. By requiring discrete data fields with standardized definitions, enumerated values for subjective assessments, and machine-readable XML formatting aligned with the mortgage industry's MISMO v3.6 standards, UAD 3.6 transforms appraisals from documents to datasets.

\begin{tcolorbox}[
    colback={rgb,255:red,218;green,224;blue,232},
    colframe={rgb,255:red,180;green,190;blue,200},
    title={\textbf{UAD 3.6 Technical Transformation}},
    fonttitle=\bfseries,
    boxrule=1pt,
    arc=4pt,
    outer arc=4pt,
    top=10pt,
    bottom=10pt,
    left=10pt,
    right=10pt
]
UAD 3.6 shifts from various reports to a single XML-based data with key changes:
\begin{itemize}
\item
  Dynamic Schema: Adapts fields based on property type
\item
  Enumerated Values: Standardized codes (C1-C6) reduce variance
\item
  MISMO v3.6 Alignment: Seamless mortgage industry integration
\item
  Granular Data: 700+ data points vs \textasciitilde150 on traditional forms
\end{itemize}
Business Impact: Enables automated QC and AI analytics but requires significant investment in software and training.
\end{tcolorbox}

The implications of this transformation extend far beyond technical formatting requirements. in \cite{williamsonMarketsHierarchiesAnalysis1975a}'s transaction cost framework, UAD 3.6 reduces asset specificity in valuation production by standardizing previously idiosyncratic processes. This standardization potentially enables new forms of market organization, from enhanced quality control through automated compliance checking to portfolio-level analytics aggregating individual valuations into market insights. The shift from narrative descriptions to structured data fields---converts tacit professional knowledge into explicit, codified information.

The specific design choices in UAD 3.6 reveal sophisticated understanding of both valuation processes and their shortcomings. Rather than simply digitizing existing forms, the standard reimagines the appraisal as a dynamic dataset that adapts to property characteristics. The replacement of multiple static forms (1004 for single-family homes, 1073 for condominiums, 1075 for exterior-only inspections) with a single dynamic Uniform Residential Appraisal Report (URAR) that adjusts its fields based on property type represents elegant solution design. When an appraiser indicates the presence of an accessory dwelling unit (ADU), relevant fields appear to capture its characteristics; for a standard single-family home without an ADU, these fields remain hidden, reducing clutter and error potential.

The standardization of subjective assessments through carefully defined enumerations addresses one of the most significant sources of inter-appraiser variability. Property condition ratings now map to specific C1-C6 categories with detailed definitions, while quality ratings use Q1-Q6 scales with clear criteria. While professional judgment still determines which category applies, the standardized framework reduces the semantic variability that plagued narrative descriptions. An appraiser can no longer describe conditions as ``average plus'' or ``good minus''---ambiguous terms that different readers interpret differently---but must select from defined categories that enable consistent interpretation and automated analysis.

\subsection{The Evolving Valuation Landscape: Alternative Methods and Market Adoption}

While UAD 3.6 represents the future standard for full appraisals, the market has simultaneously evolved to include various alternative valuation methods responding to different risk profiles and efficiency needs. Understanding this broader landscape is essential for contextualizing UAD 3.6's role and the technological challenges it presents.

In response to the time and cost associated with the traditional model, the industry has seen the emergence of alternative valuation methods. Desktop appraisals are performed entirely remotely, with the appraiser never visiting the property. Instead, they rely exclusively on third-party data sources, which offers significant speed and cost advantages but introduces a high risk of relying on outdated or inaccurate information \citep{ai2025}.

Hybrid appraisals represent a middle ground, attempting to balance cost, speed, and accuracy. In this model, the on-site data collection is often delegated to a third-party inspector---who may or may not be a licensed appraiser---while the licensed appraiser conducts the analysis and signs off on the final report remotely \citep{wire2025}. This operational split between data collection and analysis is a crucial precursor to the specialized technological solutions that are now entering the market, as it demonstrates a willingness within the industry to unbundle the traditional, monolithic role of the appraiser.

The growing use of property data collection (PDC) reports and the Uniform Property Data (UPD) standard marks another significant evolution (\autoref{tab:uad-vs-upd}). Introduced by Fannie Mae in December 2016 and Freddie Mac in June 2017, appraisal waivers have now become standard practice, with 97.9\% of GSE direct sellers issuing loans using waivers as of October 2024 \citep{jaro2025, mae2025}.

A subset of these waivers---those based on property data---relies on a UPD report instead of a full appraisal. These UPD-based waivers offer significant benefits: they reduce loan closing times by 7--10 days and save borrowers \$500--\$700 in fees. The UPD reports themselves cost roughly \$350--\$400 less than traditional appraisals and can be completed within 2--3 days \citep{jaro2025}.

\begin{table*}[htbp]
\caption{GSE Loan Appraisal Method Distribution and Trends}
\label{tab:appraisal-types}
\centering
\footnotesize
\begin{tabular}[]{@{}p{0.22\textwidth}p{0.13\textwidth}p{0.55\textwidth}@{}}\toprule\noalign{}
Metric
 & Source
 & Finding
 \\
\midrule\noalign{}
GSE loan appraisal types (Jan 2025) & \citep{center2025} & Traditional appraisals made up 82.8\% of Fannie Mae purchase loans; 1.2\% used inspection-based waivers (Value Acceptance + Property Data), and 16.1\% used traditional waivers (AEI, 2025). \\
GSE volume by method (Mar 2025) & \citep{mtgefi2025} & 84.5\% of valuation volume used traditional appraisals; 13.5\% used waivers; 2.05\% used inspection-based waivers (MtgeFi, 2025a). This marked the first month where inspection-based waivers exceeded 2\% of GSE valuations. \\
GSE volume by method (Apr 2025) & \citep{mtgefi2025a} & Traditional appraisals: 84.5\%; traditional waivers: 15.6\%; inspection-based waivers: 2.17\% (MtgeFi, 2025b). Waiver-based solutions collectively accounted for 17.8\% by volume. \\
Historical waiver share & \citep{center2024} & Waivers peaked at 44\% (Fannie) in 2021, dropped to 17--19\% in Oct 2024 (AEI, 2024). The pandemic surge demonstrated how quickly policy and market conditions can shift adoption patterns. \\
\bottomrule\noalign{}
\end{tabular}
\end{table*}

Several factors suggest continued growth in alternative valuation methods:

\begin{itemize}
\item
  Eligibility Expansion Effects: In January 2025, the FHFA expanded eligibility for inspection-based waivers to include purchase loans with loan-to-value (LTV) ratios up to 97\%. according to \cite{c2025}, this expansion could make an additional 57,000 loans eligible each month.
\item
  Hybrid Appraisals Poised for Growth: In March 2025, Fannie Mae and Freddie Mac added hybrid appraisals to their selling guides. \citep{valuation2025} reported that 94.7\% of loans originated over the past five years would have been eligible under the new hybrid policy.
\item
  Shift Toward Inspection-Based Waivers: \citep{voorheesInspectionBasedAppraisalWaivers2023} noted that GSEs increasingly favor inspection-based waivers over full appraisal waivers, signaling a long-term shift in underwriting policy.
\end{itemize}

\subsection{UAD 3.6 vs.~UPD: Understanding the Complexity Differential}

While UPD reports and UAD 3.6 reports share fundamental commonalities in their pursuit of standardized, machine-readable property data, they diverge significantly in scope, complexity, and regulatory requirements. Both frameworks represent the GSEs' broader Uniform Mortgage Data Program (UMDP) initiative to modernize mortgage data collection through structured XML formats that enable automated processing and quality control \citep{fhfa2024}.

However, UAD 3.6 reports demand substantially more comprehensive documentation, requiring detailed narrative sections, condition assessments, market analysis, and compliance with the full spectrum of USPAP standards and Fair Housing Act requirements \citep{mac2024, valuation2024}. In contrast, UPD reports focus primarily on objective property data collection---dimensions, features, and observable characteristics---without the complex analytical reasoning, comparable selection, and professional judgment components that define traditional appraisals \citep{mae2025}.

\begin{table*}[htbp]
\caption{Comparison of UAD 3.6 and UPD Requirements}
\label{tab:uad-vs-upd}
\centering
\footnotesize
\begin{tabular}[]{@{}p{0.25\textwidth}p{0.33\textwidth}p{0.33\textwidth}@{}}\toprule\noalign{}
Aspect
 & UAD 3.6
 & UPD
 \\
\midrule\noalign{}
Scope & Full appraisal with valuation opinion & Property data collection only \\
Regulatory Compliance & Full USPAP compliance required & Limited regulatory requirements \\
Analytical Requirements & Comparable selection, adjustments, market analysis & None - descriptive only \\
Professional Judgment & Extensive - condition/quality ratings, reconciliation & Minimal - objective observations \\
Narrative Sections & Multiple required (market conditions, reconciliation) & Limited or none \\
Automation Potential & Complex - requires AI reasoning capabilities & High - primarily data capture \\
\bottomrule\noalign{}
\end{tabular}
\end{table*}
This fundamental difference in cognitive complexity means that while both benefit from technological advancement, UPD workflows are inherently more amenable to automation and standardization, requiring primarily data capture and validation rather than the sophisticated analytical reasoning that UAD 3.6 demands \citep{alexandrov2023}.

Consequently, this review's focus on UAD 3.6-compliant report generation addresses a markedly different technological challenge than UPD-based workflows. While approximately 30-40\% of this review's technological foundations---particularly the physical layer discussions of 3D scanning, sensor fusion, and data acquisition---remain directly applicable to UPD report generation, the majority of the cognitive layer analysis, including agentic AI for comparable selection, market analysis, and narrative generation, is specific to the complex reasoning requirements of full appraisals \citep{amorin2024}.

The evaluation frameworks, uncertainty quantification methods, and human-in-the-loop collaboration models presented here are designed for high-stakes analytical decisions that UPD reports intentionally avoid. Therefore, while UPD reports represent a significant and growing segment of the valuation market, their streamlined, data-collection-focused workflow constitutes a fundamentally different technological problem that warrants separate analysis \citep{mba2019}. This review's concentration on UAD 3.6 addresses the more challenging frontier of automating professional analytical judgment while maintaining regulatory compliance---a capability that, once achieved, could potentially be adapted to enhance UPD workflows, but not vice versa.

\subsection{Technological Capabilities and Market Convergence}

The timing of UAD 3.6's implementation proves particularly significant as it coincides with remarkable advances in artificial intelligence capabilities directly relevant to property valuation challenges. Three categories of technological development merit detailed examination: advances in 3D scene understanding and computer vision, the emergence of large language models with sophisticated reasoning capabilities, and the evolution toward agentic AI architectures capable of autonomous task completion.

The field of 3D scene reconstruction has undergone revolutionary advancement in recent years. Classical approaches to 3D reconstruction relied on photogrammetry and structure-from-motion techniques that, while geometrically accurate, required extensive offline processing and struggled with photorealistic appearance. The introduction of Neural Radiance Fields (NeRFs) by \cite{mildenhall2020} marked a paradigm shift, demonstrating that neural networks could learn to represent complex 3D scenes from 2D images, producing photorealistic novel views. However, NeRFs suffered from extremely slow training and rendering times, limiting practical deployment.

The subsequent development of 3D Gaussian Splatting by \cite{kerbl2023} addressed these performance limitations while maintaining visual quality. By representing scenes as collections of 3D Gaussian primitives that can be efficiently rasterized, this approach achieves real-time rendering speeds while reducing training time from hours to minutes. For property documentation, this enables creation of detailed, explorable 3D models from simple video captures using consumer smartphones---democratizing technology previously requiring expensive specialized equipment.

Large language models have simultaneously demonstrated remarkable capabilities in processing unstructured text, understanding complex instructions, and generating structured outputs. The development of retrieval-augmented generation (RAG) addresses one of the key limitations of language models---their tendency to generate plausible but factually incorrect information---by grounding their outputs in retrieved documents. For appraisal applications, this enables systems that can process narrative property descriptions, extract relevant features, check against authoritative databases, and generate compliant reports while maintaining factual accuracy.

Perhaps most significantly, the emergence of agentic AI architectures capable of goal-directed behavior through planning and tool use offers frameworks for managing the full complexity of professional valuation. As described by Wang et al.~\citep{durante2024}, these systems move beyond simple input-output processing to exhibit autonomous behavior: decomposing complex tasks into subtasks, gathering information from multiple sources, using specialized tools for specific analyses, maintaining uncertainty estimates, and iterating toward solutions. For appraisal workflows, this could mean AI agents that autonomously search multiple MLS databases for comparables, analyze market trends, calculate adjustments based on paired sales analysis, and draft initial reports---all while maintaining clear audit trails of their reasoning process.

The convergence of these technological capabilities with regulatory standardization through UAD 3.6 creates what \citep{brynjolfsson2016} term ``recombinant innovation''---new possibilities emerging from novel combinations of existing technologies. The structured data requirements of UAD 3.6 provide the foundation that AI systems need to operate effectively, while AI capabilities offer solutions to the scale and consistency challenges that make UAD 3.6 implementation daunting for human appraisers alone. This convergence suggests we stand at an inflection point where fundamental transformation of valuation practice becomes not just possible but inevitable.

\section{Theoretical Framework: A Three-Layer Architecture for AI-Augmented Valuation}

\begin{figure}
    \centering
    \includegraphics[width=1\linewidth]{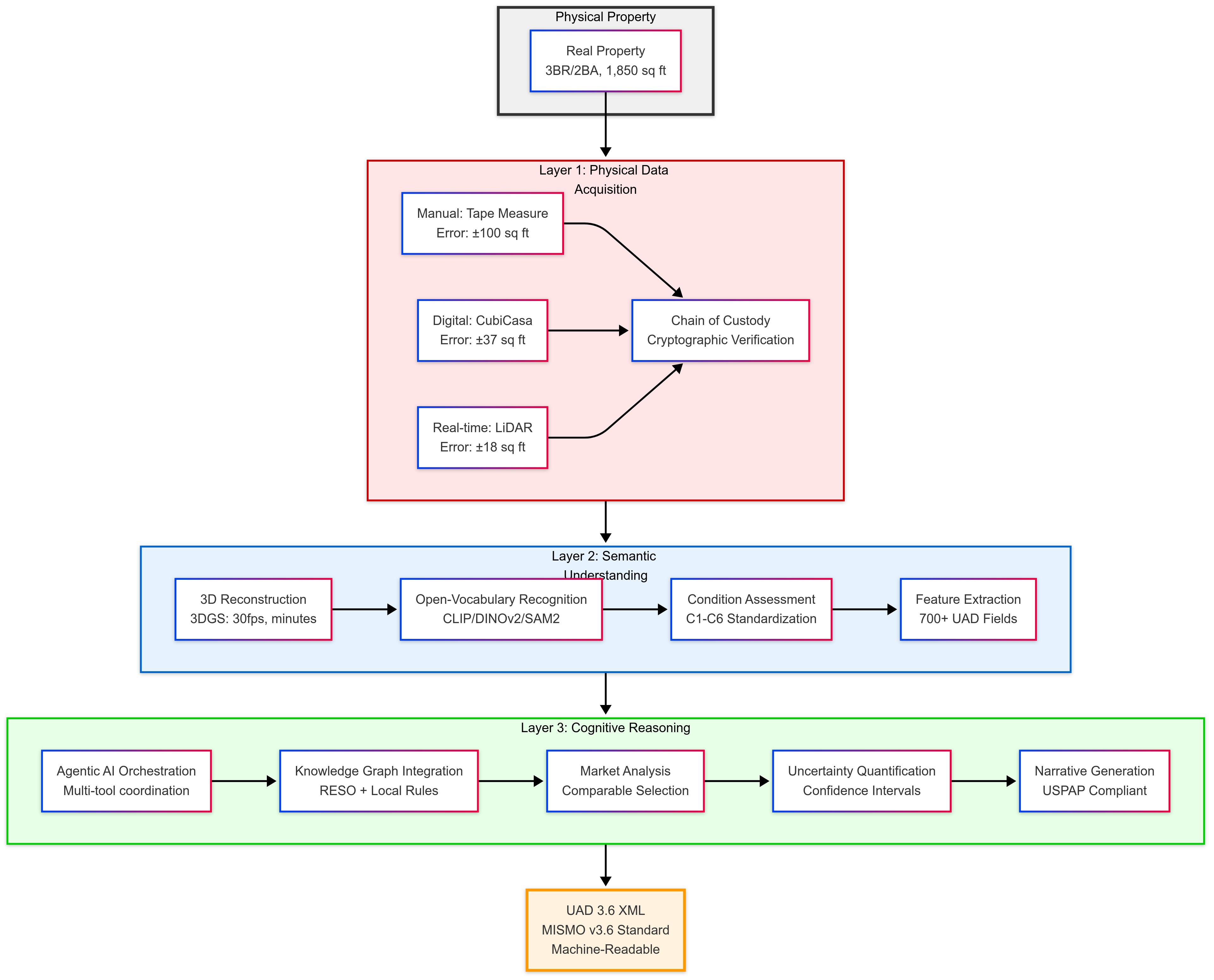}
    \caption{The three-layer architecture transforms physical properties into structured valuation data through progressive abstraction, with each layer reducing specific types of uncertainty while introducing new forms of complexity. Physical data acquisition (Layer 1) captures objective measurements with quantifiable error bounds, semantic understanding (Layer 2) interprets these measurements using visual foundation models and standardized classifications, and cognitive reasoning (Layer 3) applies professional judgment through agentic AI systems that orchestrate multiple specialized models. The framework demonstrates how the convergence of UAD 3.6's regulatory standardization with advances in AI capabilities enables fundamental transformation of valuation practice from craft-based art to data-driven science.}
    \label{fig:three-layers}
\end{figure}

Building on the institutional analysis of market failures and regulatory evolution, we now develop a comprehensive theoretical framework for understanding AI-augmented valuation systems. Our framework conceptualizes the valuation process as information transformation across three distinct but interconnected layers: physical data acquisition, semantic interpretation, and cognitive reasoning (\autoref{fig:three-layers}). This layered approach draws on established information systems architecture theory \citep{zachman1987} while incorporating insights from embodied cognition \citep{varela1991}, distributed artificial intelligence \citep{weissMultiagentSystemsModern2001a}, and professional service automation \citep{susskindFutureProfessionsHow2015a}.

\subsection{Layer One: Physical Data Acquisition and Digital Provenance}

The foundation of reliable valuation rests on accurate capture of physical property characteristics. Traditional measurement methods---tape measures, handwritten sketches, manual photography---introduce variability through both random error and systematic bias. The evolution toward automated capture technologies represents not merely efficiency improvement but fundamental enhancement in data quality, reproducibility, and verifiability. We conceptualize this evolution along a capability ladder that progresses from manual tools through digital capture to autonomous systems, with each level involving distinct trade-offs between cost, complexity, and data quality that can be understood through the lens of information economics and technology adoption theory.

\subsubsection{The Technical Evolution of Property Measurement}

At the most basic level (Level 1), manual measurement using tape measures and laser distance meters remains the predominant method among practicing appraisers. While this approach minimizes capital costs and aligns with established professional practices, it maximizes labor costs and measurement variance. Research on measurement error in appraisals suggests that manual methods introduce both random error (from imprecise measurement techniques) and systematic bias (from rounding behaviors and transcription mistakes). The absence of digital-native data requires manual re-entry into appraisal software systems, creating additional opportunities for error propagation. From an information quality perspective, manual measurement fails to create what \citep{dranove2010} identify as verifiable quality disclosure---there exists no auditable record of how measurements were obtained, making quality assessment impossible.

The transition to digital capture with offline processing (Level 2) represents an intermediate evolutionary step. Solutions like CubiCasa exemplify this approach, where users capture property videos using smartphones that are then uploaded to cloud servers for processing. Within 24 hours, the system returns professional-grade floor plans with measurements accurate to within 1-2\% and full compliance with ANSI Z765-2021 standards. The system additionally provides detailed measurement reports, gross living area calculations, and room-by-room dimensions suitable for UAD 3.6 requirements.

While dramatically improving measurement accuracy and creating digital outputs suitable for automated workflows, the temporal gap between capture and verification presents challenges---users cannot verify complete coverage during the capture process, potentially necessitating costly return visits if areas were missed or capture quality proves insufficient.

Real-time digital feedback systems (Level 3) represent the current state-of-the-art in widely accessible measurement technology. The integration of LiDAR sensors in consumer devices, particularly Apple's iPhone and iPad Pro lines since 2020, has democratized access to sophisticated 3D scanning capabilities (see \autoref{appendix:3d-scan}). Applications leveraging Apple's RoomPlan API or advanced platforms like Polycam's Space Mode generate floor plans during the scanning process, allowing users to immediately visualize coverage and ensure completeness. Canvas by Occipital, which claims to have scanned over 100 million square feet of interior space, advertises measurement accuracy within 1-2\% of professional laser measurements while reducing scanning time by 90\% compared to manual methods. These tools create immediate value by preventing the need for return visits while maintaining professional-grade accuracy suitable for lending decisions.

\subsubsection{Build vs.~Buy: Strategic Technology Decisions}

The proliferation of scanning solutions creates a critical strategic decision for organizations: whether to develop proprietary technology or partner with established providers. This decision involves complex trade-offs that extend beyond simple cost comparisons to encompass control, differentiation, and long-term competitive positioning.

The case for partnering is compelling for most organizations. Companies like CubiCasa have invested years and millions of dollars perfecting their algorithms, achieving accuracy levels that would be prohibitively expensive to replicate. Their APIs provide simple integration---submit a video, receive back standardized floor plans and measurements within 24 hours. For organizations processing fewer than 100,000 scans annually, the economics strongly favor partnership over internal development.

However, exceptions exist where vertical integration makes strategic sense. ValueMate's decision to develop proprietary scanning technology reflects their positioning as an end-to-end AI platform where scanning is merely the first step in an integrated workflow. By controlling the entire pipeline from capture to report generation, they can optimize each component for their specific use case and maintain complete control over data quality and processing speed. This approach requires significant capital investment but can create defensible competitive advantages for organizations with sufficient scale and technical capabilities.

The frontier of autonomous data acquisition (Levels 4-6) involves unmanned aerial vehicles (UAVs) and robotic systems capable of navigating properties without human guidance. While companies like Cleo Robotics have demonstrated collision-tolerant indoor drones such as the Dronut X1 Pro for industrial inspection applications, significant technical and economic barriers limit near-term deployment in residential appraisal contexts.

\subsubsection{Data Integrity and Chain of Custody}

As physical data acquisition becomes increasingly automated and potentially autonomous, establishing and maintaining data integrity becomes paramount. The concept of ``digital provenance''---a complete, tamper-evident record of data creation, processing, and transmission---addresses fundamental trust requirements in regulated financial applications.

Building trust requires implementing privacy-by-design principles from the moment of initial capture. Controlled data provenance ensures every capture creates immutable audit trails documenting when, where, and by whom data was collected. Blockchain-based approaches offer one path for creating tamper-evident records \citep{yadav2021}, though simpler cryptographic signatures may suffice for most applications \citep{damgård2015}. Companies like Truepic have demonstrated the viability of this approach with their Vision platform, which validates authenticity at the point of capture and has successfully reduced on-site inspections by 24\% in insurance applications.

Privacy considerations add another layer of complexity to physical data acquisition in residential contexts. Unlike traditional photography that captures specific views, comprehensive 3D scans create detailed permanent records of private living spaces, personal possessions, and lifestyle patterns. This data richness, while valuable for appraisal accuracy and fraud prevention, raises significant privacy concerns that must be addressed proactively \citep{quadri2025}.

\begin{tcolorbox}[
    colback={rgb,255:red,218;green,224;blue,232},
    colframe={rgb,255:red,180;green,190;blue,200},
    title={\textbf{Privacy-Preserving 3D Capture}},
    fonttitle=\bfseries,
    boxrule=1pt,
    arc=4pt,
    outer arc=4pt,
    top=10pt,
    bottom=10pt,
    left=10pt,
    right=10pt
]
\begin{itemize}
\item Technical: Selective capture, edge blur processing, semantic filtering, encrypted storage.
\item Regulatory: GDPR deletion rights, CCPA disclosure requirements, fair lending restrictions.
\item Implementation: Auto-exclude photos, personal documents, children's rooms beyond dimensions, valuable items.
\item Business Impact: Privacy-first design builds trust but adds \textasciitilde15-20\% system complexity.
\end{itemize}
\end{tcolorbox}

\subsection{Layer Two: Perception and Semantic Understanding}

Converting raw sensor data into semantically meaningful property representations constitutes the second critical transformation in our framework. This process involves both geometric reconstruction---creating accurate 3D models of physical spaces---and semantic interpretation---understanding what objects and spaces represent in terms of property features and conditions. The theoretical challenges at this layer mirror fundamental problems in artificial intelligence regarding symbol grounding, scene understanding, and knowledge representation.

\subsubsection{The Evolution of 3D Reconstruction: From Geometry to Neural Representations}

The field of 3D scene reconstruction has undergone revolutionary advancement that fundamentally changes what is possible in property documentation. Classical approaches to 3D reconstruction relied on explicit geometric primitives---point clouds from photogrammetry or LiDAR scanning, polygon meshes from surface reconstruction algorithms. Software systems like COLMAP \citep{schonberger2016} could process collections of photographs to produce sparse point clouds and camera poses through Structure-from-Motion (SfM), followed by dense reconstruction using Multi-View Stereo (MVS) techniques. While geometrically accurate and well-understood from decades of computer vision research, these representations struggled with several limitations: inability to capture view-dependent effects like reflections and transparency, requirement for extensive manual cleanup in complex scenes, computationally expensive processing, and lack of photorealism in rendered views.

The introduction of Neural Radiance Fields (NeRFs) by \cite{mildenhall2020} marked a paradigm shift in 3D representation. Rather than explicitly modeling geometry, NeRFs encode entire scenes within neural network weights, learning to map 3D coordinates and viewing directions to color and density values. This implicit representation achieves unprecedented photorealism in novel view synthesis, accurately capturing complex lighting effects, reflections, and transparency that classical methods struggle with. For property documentation, this means the ability to create virtual tours that are virtually indistinguishable from real photography, capturing subtle material properties and lighting conditions that affect perceived property quality.

However, NeRFs' computational requirements severely limited practical deployment. Training a single scene required hours on high-end GPUs, while rendering each frame took seconds---making interactive viewing impossible. The implicit representation also complicated extracting geometric measurements needed for appraisals. These limitations drove rapid innovation in neural scene representation.

The development of 3D Gaussian Splatting by \cite{kerbl2023} elegantly addressed these performance limitations while maintaining visual quality. By representing scenes as collections of 3D Gaussian primitives---each parameterized by position, covariance matrix, opacity, and view-dependent color---the method achieves the best of both worlds. The explicit primitive representation enables efficient rasterization-based rendering at 30+ frames per second, while the continuous Gaussian representation maintains the ability to capture soft boundaries and complex appearance. Training time reduces from hours to minutes, making the technology practical for routine use.

Recent advances push these capabilities even further. Feed-forward methods eliminate scene-specific optimization entirely---models like InstantSplat \citep{fan2025} can generate 3D Gaussian representations from sparse images in seconds rather than minutes. Compression techniques achieve 10x file size reduction while maintaining visual quality, addressing storage and transmission constraints for practical deployment. The open-source SPZ format developed by Niantic Labs demonstrates approximately 10x compression compared to standard PLY files with virtually no perceptible quality loss.

\subsubsection{Critical Analysis: Why 3D Gaussian Splatting Struggles with Buildings}

Despite 3D Gaussian Splatting's impressive performance for natural scenes, fundamental limitations emerge when applying it to indoor environments dominated by architectural elements. Understanding these limitations has driven extensive research into alternative representations better suited to the geometric primitives of built environments.

The core challenge stems from a fundamental representational mismatch. Gaussian primitives are essentially fuzzy 3D ellipsoids with smooth, continuous density falloff---ideal for representing organic shapes, vegetation, and textured surfaces. However, indoor spaces consist primarily of sharp-edged planar surfaces: walls, floors, ceilings, doorways, and windows. Representing a perfectly flat wall requires hundreds or thousands of overlapping Gaussians, creating both computational inefficiency and visual artifacts. The accumulation of many soft primitives attempting to approximate a hard surface results in a subtle but perceptible ``cloudiness'' or texture where none should exist---what practitioners colloquially describe as the ``cotton ball effect.''

Thin structures present another fundamental challenge for volumetric Gaussian representations. Architectural elements like table legs, chair backs, stair railings, window frames, and door hardware are essentially one-dimensional or two-dimensional structures in a 3D world. Volume-occupying Gaussians cannot represent such structures without artificially inflating their apparent thickness. A chair leg that should appear as a crisp 1-inch diameter cylinder instead renders as a fuzzy 2-3 inch blob, fundamentally altering the furniture's appearance and potentially affecting quality assessments. For appraisals where furniture quality and architectural details contribute to property value, such distortions prove problematic.

Reflective and transparent surfaces---ubiquitous in modern construction---create particular difficulties for standard Gaussian representations. The fundamental issue, which researchers term the ``transparency-depth dilemma'' \citep{li2025}, arises because the representation cannot simultaneously model a surface as transparent (to show what lies behind) and opaque (to show reflections). This leads to systematic failures in reconstructing windows, glass doors, mirrors, and other reflective surfaces. Windows may disappear entirely, create floating artifacts where reflections should appear, or be incorrectly positioned in depth. For properties where natural light, views, and modern glass features significantly impact value, accurate reconstruction of these elements proves essential.

\subsubsection{Alternative Primitives: The Research Response}

These limitations have catalyzed an explosion of research into alternative primitive representations, each attempting to better match the geometric characteristics of built environments  (see \autoref{appendix:3d-repr}.

The proliferation of alternative representations suggests a fundamental insight: no single primitive type can optimally handle all aspects of complex indoor environments. This aligns with the ``no free lunch'' theorem in machine learning \citep{wolpert1997}, implying that specialized representations for different scene elements may prove necessary. Future systems will likely employ adaptive approaches that automatically select appropriate representations based on local geometry: planar primitives for walls and floors, specialized transparent surface handlers for windows, thin structure representations for furniture details, and standard Gaussians for organic clutter and vegetation.

\begin{tcolorbox}[
colback={rgb,255:red,218;green,224;blue,232},
colframe={rgb,255:red,180;green,190;blue,200},
title={\textbf{3DGS, NeRFs, and Alternatives: The Reality Check}},
fonttitle=\bfseries,
boxrule=1pt,
arc=4pt,
outer arc=4pt,
top=10pt,
bottom=10pt,
left=10pt,
right=10pt
]
\begin{itemize}
\item \textbf{Winner:} 3DGS dominates (30+ fps, minutes to train) despite indoor flaws—cotton ball walls, thickened edges, glass failures.
\item \textbf{NeRFs:} Beautiful but dead for proptech—hours to train, seconds per frame.
\item \textbf{Indoor Problem:} Buildings need sharp edges, not fuzzy clouds. Alternatives (planar/convex primitives) use 45-82\% fewer parameters with better accuracy—but lack adoption.
\item \textbf{Market Reality:} Network effects beat technical superiority. 3DGS has ecosystem lock-in (tools, tutorials). PropTech wants "good enough" at scale.
\item \textbf{Impact:} 3DGS artifacts acceptable for UAD 3.6—appraisers verify dimensions, not aesthetics. Future: hybrid approaches for critical measurements.
\end{itemize}
\end{tcolorbox}

\subsubsection{Semantic Understanding and Open-World Recognition}

Beyond geometric reconstruction lies the critical challenge of semantic understanding---not merely capturing what spaces look like but comprehending what objects are, their conditions, relationships, and implications for property value. Traditional computer vision approaches to scene understanding relied on closed-vocabulary systems trained to recognize predefined object categories. A system might be trained to identify kitchens, bedrooms, bathrooms, and standard appliances. However, such closed-world approaches fail catastrophically when encountering the rich diversity of real-world properties: custom architectural features, unusual room configurations, specialized equipment, or emerging home technologies like solar panels and EV charging stations.

The integration of Visual Foundation Models (VFMs) represents a paradigm shift toward open-vocabulary scene understanding. Models like CLIP (Contrastive Language-Image Pre-training), developed by \cite{radford2021}, learn rich visual-linguistic representations by training on hundreds of millions of image-text pairs from the internet. Rather than learning fixed categories, these models learn to align visual and textual concepts in a shared embedding space. DINOv2 \citep{oquab2024} provides powerful visual features through self-supervised learning without requiring labeled data. SAM2 (Segment Anything Model 2) \citep{faigle2024} enables precise object segmentation without predefined categories, identifying distinct objects and their boundaries from visual properties alone.

These foundation models transform property analysis capabilities. Instead of systems limited to recognizing ``kitchen appliances'' from a fixed list, open-vocabulary models can identify and describe ``Sub-Zero refrigerator with French doors showing minor scratches on the lower panel'' from natural language queries. They can distinguish between ``engineered hardwood with hand-scraped texture'' and ``luxury vinyl plank flooring with wood-grain pattern''---distinctions that significantly impact property value but would be impossible for traditional closed-vocabulary systems.

\begin{tcolorbox}[
    colback={rgb,255:red,218;green,224;blue,232},
    colframe={rgb,255:red,180;green,190;blue,200},
    title={\textbf{Open-Vocabulary Scene Understanding}},
    fonttitle=\bfseries,
    boxrule=1pt,
    arc=4pt,
    outer arc=4pt,
    top=10pt,
    bottom=10pt,
    left=10pt,
    right=10pt
]
\begin{itemize}
\item Traditional: Fixed categories fail on novel items (``wine refrigerator'').
\item Modern: Natural language understanding handles ``home theater with acoustic panels.''
\item Key VLMs: CLIP (image-text alignment), DINOv2 (visual features), SAM2 (segmentation).
\item Business Impact: Identifies value-affecting features beyond standard categories.
\end{itemize}
\end{tcolorbox}

For practical deployment in appraisal workflows, systems like OpenScene \citep{peng2023} demonstrate how to process point clouds with open-vocabulary capabilities. These approaches enable sophisticated analyses critical for valuation:

\begin{itemize}
\item
  Natural language queries for measurement: ``Calculate the square footage of all carpeted areas on the second floor''
\item
  Condition assessment from visual cues: ``Identify all surfaces showing wear, deterioration, or damage''
\item
  Feature extraction and cataloging: ``List all kitchen appliances with their approximate age and condition ratings''
\item
  Spatial relationship understanding: ``Find all bedrooms with attached bathrooms'' or ``Identify rooms with direct outdoor access''
\end{itemize}

The implications extend beyond simple object recognition to understanding scene affordances---the possibilities for action that environments offer. Modern AI systems can recognize that a finished basement could serve as additional bedrooms (subject to egress requirements), understand that dated kitchen cabinets represent a value-limiting factor suggesting renovation potential, or identify that a detached structure could function as an accessory dwelling unit (ADU) adding significant value in markets with ADU-friendly zoning.

This semantic understanding bridges the gap between raw sensor data and professional judgment. By automatically identifying and cataloging property features, assessing conditions, and recognizing value-relevant patterns, these systems transform time-consuming manual inspection processes into efficient, consistent, and comprehensive analyses. However, the semantic layer serves not to replace professional judgment but to augment it---providing appraisers with rich, structured information as the foundation for their expert evaluation.

\subsection{Layer Three: Cognitive Reasoning and Knowledge Integration}

The highest layer of our framework addresses the transformation from structured observations and semantic understanding into professional judgments and compliant valuation reports. This cognitive layer requires capabilities that extend far beyond pattern matching or information retrieval, encompassing causal reasoning about market dynamics, integration of diverse information sources, uncertainty quantification for risk assessment, regulatory compliance verification, and narrative generation that satisfies both technical and legal requirements. We conceptualize this layer through the lens of distributed cognition \citep{hollan2000}, where intelligence emerges from the interaction of human expertise, artificial intelligence capabilities, and institutional knowledge encoded in regulations and professional standards.

\subsubsection{Agentic AI Architectures for Professional Reasoning}

The complexity of professional valuation workflows makes them ideal candidates for agentic AI architectures---systems that move beyond simple input-output processing to exhibit goal-directed autonomous behavior. as defined by \cite{acharya2025}, agentic systems possess several key characteristics that align with appraisal requirements: the ability to perceive their environment (property data, market conditions), maintain goals (producing accurate, compliant valuations), plan sequences of actions to achieve those goals, and utilize tools and external resources as needed.

For appraisal workflows, agentic architectures enable systems that mirror the iterative, investigative nature of professional practice. Rather than attempting to generate valuations in a single pass, these systems decompose complex valuation tasks into manageable subtasks through hierarchical planning. When tasked with generating a UAD 3.6 compliant appraisal, an agentic system would systematically verify property identification and legal description, validate physical characteristics against multiple sources, identify and analyze comparable sales using multiple search strategies, calculate adjustments using paired sales analysis and market extraction, assess market conditions through trend analysis, generate narrative sections with appropriate supporting evidence, and perform quality checks for internal consistency and regulatory compliance.

The autonomous information gathering capabilities of modern agentic systems demonstrate sophisticated tool use that transforms the appraisal workflow \citep{qin2023}. These systems search multiple MLS databases using different query strategies to find truly comparable properties, access public records for ownership history, tax assessments, and permit data, utilize mapping APIs for location analysis and neighborhood delineation, invoke specialized calculators for depreciation schedules or capitalization rate derivation, call image analysis models for detailed condition assessment, and connect to economic data providers for market trend analysis. This autonomous capability fundamentally changes the nature of professional work, shifting human effort from data gathering to validation and interpretation.

Real-world appraisal scenarios rarely provide complete, consistent information, requiring agentic systems to reason under uncertainty and incomplete information. These systems must recognize when data conflicts exist between sources, identify when critical information is missing, express appropriate levels of confidence in their conclusions, and determine when human expertise is required. This capability proves essential for maintaining professional standards where acknowledging limitations is both ethically required and legally prudent. The ability to gracefully handle incomplete information distinguishes professional-grade AI systems from simpler automation tools.

Unlike static models that remain fixed after training, agentic systems can learn and adapt through experience. They develop understanding of local market nuances not captured in training data, such as learning that waterfront premiums vary seasonally in resort markets or that certain neighborhood boundaries create sharp value discontinuities. They adapt to individual appraiser preferences for report formatting and terminology while maintaining regulatory compliance, recognize patterns in revision requests to prevent recurring issues, and update their knowledge bases with new regulations, market conditions, and professional guidance. This continuous learning capability ensures systems remain current and relevant in dynamic market conditions.

The architectural advantages of agentic approaches for professional services prove substantial. Transparency emerges through observable actions where every database query, calculation, and intermediate decision can be logged and audited, providing clear traces of reasoning that satisfy regulatory requirements for documented support. Modularity enables targeted improvements, allowing the comparable selection module to be updated without affecting report generation, facilitating rapid adaptation to market changes or regulatory updates. Human-in-the-loop integration occurs naturally at decision points rather than just final approval, enabling nuanced collaboration where human expertise guides AI processing. When optimal data sources are unavailable, graceful degradation ensures system robustness through pursuit of alternative strategies rather than complete failure.

\begin{tcolorbox}[
    colback={rgb,255:red,218;green,224;blue,232},
    colframe={rgb,255:red,180;green,190;blue,200},
    title={\textbf{Agentic AI for Appraisal Workflows}},
    fonttitle=\bfseries,
    boxrule=1pt,
    arc=4pt,
    outer arc=4pt,
    top=10pt,
    bottom=10pt,
    left=10pt,
    right=10pt
]
\begin{itemize}
\item Traditional AI: User provides all data $\rightarrow$ single-shot output.
\item Agentic AI: Autonomously searches sources, iteratively refines, adapts workflow.
\item Example: Receives address $\rightarrow$ queries MLS/tax/permits $\rightarrow$ resolves conflicts $\rightarrow$ finds comparables $\rightarrow$ flags human review needs.
\item Business Impact: 60-70\% reduction in data gathering while improving completeness.
\end{itemize}
\end{tcolorbox}

\subsubsection{Knowledge Infrastructure as Institutional Memory}

\begin{figure}
    \centering
    \includegraphics[width=1\linewidth]{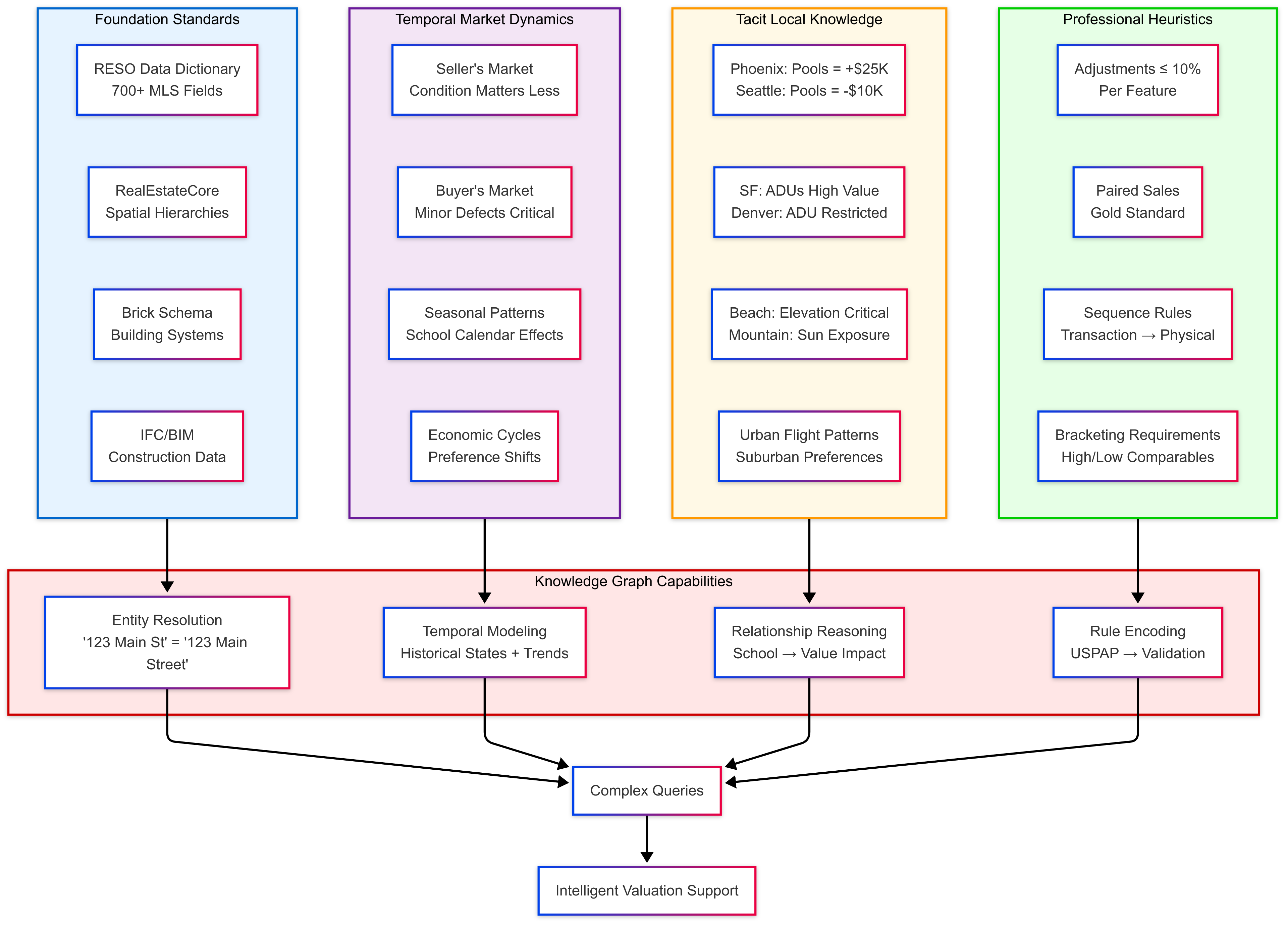}
    \caption{The Real Estate Knowledge Graph integrates formal standards like RESO's 700+ fields with tacit local knowledge accumulated through decades of professional practice, capturing not just facts but the relationships, rules, and temporal patterns that distinguish expert valuation from mechanical calculation. Critical capabilities include entity resolution to handle the pervasive problem of duplicate property identification across systems, temporal modeling to track how properties and markets evolve, relationship reasoning to understand complex value influences, and rule encoding to ensure regulatory compliance. This knowledge infrastructure serves as institutional memory, preserving expertise that might otherwise be lost as the profession undergoes demographic transition, while enabling AI systems to reason with the nuance and context-awareness that characterizes professional judgment.}
    \label{fig:kg}
\end{figure}

While large language models trained on internet-scale text demonstrate impressive general capabilities, their application to specialized professional domains requires grounding in structured domain knowledge. This need drives the development of comprehensive Real Estate Knowledge Graphs (KGs) that serve as institutional memory---capturing not just facts but the relationships, rules, and patterns that define professional practice (\autoref{fig:kg}). Building effective knowledge infrastructure requires synthesis of multiple perspectives from formal ontology design in information science, domain modeling from real estate practice, and knowledge engineering from artificial intelligence.

The foundation of knowledge infrastructure begins with established real estate data standards that provide vocabulary and structure (\autoref{tab:ontologies}). The Real Estate Standards Organization (RESO) Data Dictionary encompasses over 700 standardized fields ensuring MLS compatibility across North America. RealEstateCore provides ontological modeling of spatial hierarchies from buildings to floors to individual rooms, essential for understanding property relationships. Brick Schema captures the complex relationships between building systems, equipment, and their maintenance implications for property value. Industry Foundation Classes (IFC) enable integration with Building Information Modeling (BIM) data, facilitating cost approach calculations and renovation assessments.

\begin{table*}[htbp]
\caption{Comparison of Real Estate Ontologies/Standards}
\label{tab:ontologies}
\centering
\footnotesize
\begin{tabular}[]{@{}p{0.13\textwidth}p{0.19\textwidth}p{0.20\textwidth}p{0.18\textwidth}p{0.20\textwidth}@{}}\toprule\noalign{}
Feature
 & RESO Data Dictionary
 & RealEstateCore (REC)
 & Brick Schema
 & Industry Foundation Classes (IFC)
 \\
\midrule\noalign{}
Primary Use Case & Transactional Data Exchange (MLS) & Smart Building/Digital Twin Management & Building Systems Modeling & Building Information Modeling (BIM) \\
Core Data Model & Standardized flat fields and enumerations & RDF/SHACL/DTDL Graph Ontology & RDF/OWL Graph Ontology & EXPRESS data modeling language \\
Scope & Property listing, sale, and agent data & Building lifecycle, systems, sensors, spatial topology & Building systems, points, and equipment & Detailed building geometry and construction data \\
Key Entities & Property, Member, Office, Media & Building, Space, Asset, Point, Agent & Equipment, Location, Point, Tag & IfcProduct, IfcProject, IfcActor \\
Relationship Handling & Implicit relationships via foreign keys & Explicitly defined semantic relationships (hasPart, locatedIn) & Explicit semantic relationships (hasLocation, hasPoint) & Complex object-oriented relationships \\
Extensibility & Local custom fields & Ontological extension and alignment & Designed for extension & Extensible through property sets \\
\bottomrule\noalign{}
\end{tabular}
\end{table*}

\begin{tcolorbox}[
    colback={rgb,255:red,218;green,224;blue,232},
    colframe={rgb,255:red,180;green,190;blue,200},
    title={\textbf{Real Estate Knowledge Graphs}},
    fonttitle=\bfseries,
    boxrule=1pt,
    arc=4pt,
    outer arc=4pt,
    top=10pt,
    bottom=10pt,
    left=10pt,
    right=10pt
]
\begin{itemize}
\item Content: Entities, relationships, attributes, and domain rules.
\item Standards: RESO (700+ MLS fields), RealEstateCore (spatial), Brick Schema (systems), IFC (BIM).
\item Value: Consistent terminology, complex queries, local market rules, institutional knowledge preservation.
\end{itemize}
\end{tcolorbox}

However, effective knowledge graphs extend beyond standards to capture the tacit knowledge that distinguishes professional practice. Local market rules and conventions vary dramatically by geography. In Phoenix, pools represent valuable amenities that increase property value, while in Seattle, they may detract due to maintenance concerns and limited use seasons. San Francisco's treatment of in-law units differs substantially from Denver's approach due to different zoning philosophies and housing pressures. Beach communities value elevation as protection against flood risk, while mountain communities prioritize southern exposure for solar gain and snow management. These location-specific patterns, accumulated through decades of professional experience, must be systematically encoded for accurate valuation.

Temporal market dynamics introduce another layer of complexity that knowledge graphs must capture. During seller's markets characterized by low inventory and high demand, property condition matters less as buyers compete for any available inventory. Conversely, in buyer's markets with abundant supply, minor defects can significantly impact value as purchasers become more selective. Seasonal patterns affect different property types in distinct ways---vacation homes experience peak demand in spring as buyers prepare for summer use, while family homes align with school calendars, showing increased activity in spring for summer moves. Economic cycles create shifting preferences between urban and suburban locations, with flight to suburbs during economic uncertainty and urban renaissance during growth periods.

Professional heuristics and adjustment guidelines developed over decades of practice provide essential guardrails for AI systems. Industry conventions dictate that adjustments typically shouldn't exceed 10\% for any single feature, preventing unrealistic valuations based on minor differences. Paired sales analysis remains the gold standard for deriving adjustments, providing market-based evidence for value differences. When paired sales are unavailable, practitioners employ cost-based derivations using replacement cost data or income-based approaches using rent multipliers. The sequence of adjustments follows established patterns, with transactional adjustments applied before physical characteristics to maintain logical consistency. These heuristics, while sometimes appearing arbitrary to outsiders, embody collective professional wisdom about market behavior and valuation reliability.

The knowledge graph serves multiple critical functions in the cognitive architecture through sophisticated capabilities extending beyond simple data storage. Entity resolution and disambiguation addresses the pervasive problem of duplicate and inconsistent property identification across systems. The same property might appear as ``123 Main St,'' ``123 Main Street,'' ``123 Main St Unit A,'' or with various abbreviations and formats. The knowledge graph maintains canonical entities with comprehensive alias lists, postal standardization rules, and probabilistic matching algorithms ensuring consistent property identification regardless of source variations.

Temporal modeling enables sophisticated analysis of market evolution through timestamped facts and relationship histories. Properties and markets change continuously through renovations that alter characteristics, zoning changes that affect use potential, and market conditions that fluctuate with economic cycles. Temporal knowledge graphs enable queries examining historical conditions (``What was this property's condition rating in 2019?'') or trend analysis (``How have prices for similar properties changed over the past 5 years?''). This temporal awareness proves essential for selecting truly comparable sales from similar market conditions and understanding value trends that inform current valuations.

Relationship reasoning captures the complex web of factors affecting property value through multi-dimensional analysis. The knowledge graph models proximity relationships including walkability scores, school attendance zones, and distance to amenities that significantly impact value. It captures influence patterns showing how commercial development affects nearby residential values or how transportation improvements ripple through neighborhoods. Comparability factors extend beyond simple feature matching to include market positioning and buyer profiles, recognizing that properties may be comparable despite surface differences if they appeal to similar purchasers. Market segmentation acknowledges that luxury and entry-level homes behave according to different dynamics, requiring separate analysis frameworks.

Rule encoding transforms regulatory requirements and professional standards into computationally actionable form, ensuring AI systems operate within established bounds. USPAP requirements for documentation and analysis scope become verification checklists that systems must satisfy before completing valuations. Fannie Mae guidelines for acceptable adjustment ranges become validation rules preventing unrealistic valuations. Local zoning regulations affecting highest and best use become automated checks ensuring legal compliance. This systematic encoding of rules and requirements ensures AI systems not only produce accurate valuations but also maintain professional and regulatory compliance throughout the process.

\subsubsection{Advanced Reasoning Capabilities}

The evolution of agentic AI systems has introduced sophisticated reasoning capabilities that extend far beyond simple pattern matching or rule application. These advanced capabilities enable nuanced professional judgment that approaches human expertise while maintaining the consistency and scale advantages of automated systems.

Market trend analysis and prediction represents a critical area where AI reasoning capabilities excel through integration of diverse data sources and sophisticated pattern recognition. By analyzing vast datasets encompassing transaction histories, economic indicators, demographic shifts, and even social media sentiment, AI systems identify subtle patterns invisible to human analysts. These systems detect early signals of neighborhood gentrification through combinations of business permit applications, demographic changes, and social media activity patterns. They recognize seasonal variations that differ by property type and price range, understanding that luxury vacation homes follow different seasonal patterns than entry-level family residences. The identification of correlations between seemingly unrelated factors, such as coffee shop density serving as a leading indicator of property appreciation, demonstrates the system's ability to discover non-obvious relationships. Rather than relying on simple linear extrapolation, these systems project future values using complex multivariate models that account for economic cycles, demographic trends, and policy changes.

Comparable selection optimization showcases how AI can enhance one of appraisal's most judgment-intensive tasks through sophisticated multi-criteria analysis. Traditional approaches rely on simple filters based on proximity, size, and age combined with appraiser intuition. AI systems consider hundreds of variables simultaneously, dynamically weighting their importance based on local market dynamics and property-specific factors. In urban markets, the system might recognize that walkability score and public transit access matter more than lot size, while in suburban areas, school quality and neighborhood amenities dominate other factors. The system identifies non-obvious comparables that share important but subtle characteristics, perhaps matching homes with similar renovation quality despite different construction dates, or properties that appeal to the same buyer demographic despite variations in style or size. This nuanced matching goes beyond mechanical filtering to understand what truly makes properties comparable in specific market contexts.

Adjustment calculation and validation benefits from AI's ability to perform sophisticated statistical analysis across large datasets, moving beyond simple rules of thumb to market-based derivation. The system automatically identifies truly comparable pairs for paired sales analysis, controlling for multiple variables to isolate the impact of specific features. Regression analysis simultaneously accounts for numerous property characteristics, market conditions, and temporal factors to derive accurate adjustments. Depreciation modeling based on actual market behavior rather than theoretical curves captures how different property components lose value over time in specific markets. Sensitivity analysis reveals how adjustments interact, ensuring that the cumulative impact of multiple adjustments remains reasonable. The system validates proposed adjustments against market data, flagging when suggested values fall outside reasonable ranges based on historical patterns and current market conditions.

Risk assessment and uncertainty quantification represents perhaps the most valuable contribution of advanced AI reasoning to professional valuation practice. Traditional appraisals provide point estimates with little indication of confidence levels or risk factors. AI systems quantify uncertainty arising from multiple sources throughout the valuation process. Market volatility in rapidly changing areas creates valuation uncertainty that the system captures through confidence intervals. Data quality issues arise when information sources conflict, requiring the system to assess source reliability and propagate uncertainty through calculations. Comparable scarcity in unique property situations leads to wider confidence intervals reflecting the increased difficulty of accurate valuation. Model uncertainty emerges when dealing with unusual properties that fall outside normal parameters, prompting the system to flag these cases for additional review. This comprehensive uncertainty quantification enables more sophisticated risk-based decision-making by lenders and provides transparent communication of valuation confidence to all stakeholders.

The integration of these advanced reasoning capabilities transforms property valuation from a largely manual, inconsistent process to a sophisticated analytical framework that combines the best of human expertise with AI's computational power. The result is not replacement of professional judgment but its augmentation, enabling appraisers to focus on complex cases requiring human insight while AI handles routine analysis and ensures comprehensive, consistent evaluation across all properties.

\section{Trust, Accountability, and Systemic Risk Management}

The application of artificial intelligence to property valuation raises fundamental questions about trust, accountability, and systemic risk that extend far beyond technical performance metrics. Unlike consumer applications where errors result in minor inconvenience, property valuations directly impact mortgage underwriting decisions, homeowner wealth, and financial system stability. This section examines these critical concerns through multiple theoretical lenses, integrating perspectives from law and economics, algorithmic accountability literature, behavioral finance, and financial stability frameworks.

\subsection{Regulatory Compliance and the Explainability Imperative}

The regulatory environment for property valuation reflects broader tensions in algorithmic accountability---legal frameworks developed for human decision-making struggle to accommodate autonomous systems. This creates what \citep{krollAccountableAlgorithms2016a} term the ``accountability gap,'' where traditional mechanisms of professional oversight and liability become ambiguous when algorithms contribute to decisions. For property valuation, this gap proves particularly problematic given the web of federal regulations, state licensing requirements, and professional standards that govern practice.

The Uniform Standards of Professional Appraisal Practice (USPAP) exemplifies these challenges. USPAP's Standards Rule 1-1 requires that appraisers ``be aware of, understand, and correctly employ those recognized methods and techniques that are necessary to produce a credible appraisal'' \citep{foundation2024}. This requirement assumes human judgment in selecting and applying appropriate methods. When AI systems make these methodological choices---selecting comparables, determining adjustment factors, weighting different approaches to value---questions arise about competency, supervision, and ultimate responsibility.

The competency provision proves particularly challenging. USPAP requires appraisers to have competency in the property type, geographic area, and analytical methods employed. But what constitutes competency when using AI tools? Must appraisers understand the mathematical foundations of neural networks? Can they rely on AI systems for market areas where they lack personal experience? The Appraisal Standards Board has yet to provide definitive guidance, leaving practitioners to navigate uncertainty.

Recent regulatory developments attempt to address these concerns while maintaining flexibility for innovation. The Consumer Financial Protection Bureau's position that there is ``no black box exemption'' from fair lending laws establishes clear expectations \citep{cfpb2022}. Financial institutions cannot escape liability by claiming algorithmic complexity prevents explanation. This aligns with broader international trends toward algorithmic transparency, including the EU's General Data Protection Regulation provisions on automated decision-making and proposed AI Act requirements for high-risk applications.

The technical challenge of explainability in neural systems intersects with fundamental epistemological questions about knowledge and justification. Drawing on philosophical work on testimonial knowledge \citep{coady1992}, accepting AI-generated valuations requires understanding not just what the system concluded but how it reached those conclusions. This epistemic requirement drives adoption of explainable AI (XAI) techniques that attempt to illuminate the ``black box'' of neural processing.

Current approaches to explainability in machine learning offer partial solutions with different tradeoffs:

SHAP (SHapley Additive exPlanations) values, developed by \cite{lundberg2017}, provide game-theoretic attribution of predictions to input features. For property valuation, SHAP can quantify how each characteristic contributes to final value estimates. A typical SHAP analysis might reveal:

\begin{itemize}
\item
  Location factors: +\$50,000 (proximity to top-rated elementary school, walkable neighborhood)
\item
  Size premium: +\$75,000 (500 sq ft above neighborhood average)
\item
  Condition adjustment: -\$20,000 (original kitchen and bathrooms indicating deferred updates)
\item
  Recent improvements: +\$15,000 (new roof installed 2023, HVAC updated 2022)
\end{itemize}

This additive explanation model aligns naturally with traditional appraisal methods where adjustments are explicitly calculated and summed, making SHAP particularly intuitive for practitioners to validate and regulators to review.

LIME (Local Interpretable Model-agnostic Explanations), proposed by \cite{ribeiro2016}, creates simplified local linear approximations of complex models around specific predictions. For a particular property, LIME might reveal that within the local neighborhood of similar 3-bedroom homes built in the 1990s, each additional bathroom adds \$15,000, while a pool adds only \$5,000 due to maintenance concerns. This local interpretation proves valuable for explaining individual decisions while acknowledging that relationships may differ in other contexts.

Counterfactual explanations, as formalized by \cite{wachter2017}, show what would need to change for different valuation outcomes. These prove particularly valuable for fair lending compliance and consumer communication: ``The property would appraise \$50,000 higher with an updated kitchen and primary bathroom renovation'' or ``Moving the property 0.5 miles west would place it in a different neighborhood, increasing value by \$35,000.'' Such explanations provide actionable insights while revealing the factors driving valuation.

These counterfactual approaches share conceptual similarities with traditional paired sales analysis (\autoref{tab:paired-counterfactual})---a cornerstone of appraisal practice. Both methods seek to isolate the impact of specific features on property value, but they operate through fundamentally different mechanisms. Traditional paired sales analysis relies on empirical market evidence, with appraisers identifying actual transactions of nearly identical properties that differ in only one or two characteristics. This provides direct, market-based evidence of feature values. Counterfactual explanations, in contrast, reveal how AI models internally represent feature importance by showing the minimal changes needed to alter a prediction. While paired sales ground adjustments in observable market behavior, counterfactuals illuminate the decision logic within complex machine learning models, making their reasoning transparent and auditable.

\begin{table*}[htbp]
\caption{Comparison of Paired Data Analysis and Counterfactual Explanations in Property Valuation}
\label{tab:paired-counterfactual}
\centering
\footnotesize
\begin{tabular}[]{@{}p{0.15\textwidth}p{0.38\textwidth}p{0.40\textwidth}@{}}\toprule\noalign{}
Feature
 & Paired Data Analysis (Traditional Appraisal)
 & Counterfactual Examples (XAI for AVMs)
 \\
\midrule\noalign{}
Goal & To determine the actual market value contribution of a specific feature & To explain why a model made a specific prediction by showing how to change it \\
Methodology & Compares two real historical sales that are nearly identical & Generates hypothetical changes to input features to alter a model's prediction \\
Data Source & Real, observable sales data from the market & The input features to a predictive model; the model itself generates the counterfactual \\
Output & A dollar or percentage adjustment representing the feature's contributory value & An ``if-then'' statement showing minimal feature changes for a different prediction \\
Human Role & Appraiser selects comparables, performs adjustments, applies judgment & Algorithm generates explanations; human interprets the explanation \\
Transparency & Inherently transparent; logic is based on observable market behavior & Aims to bring transparency to complex, often opaque, AI models \\
Complexity Handled & Best for isolating single or very few variables & Can explain predictions from highly complex, non-linear models \\
Actionability & Provides a direct market-based value & Provides actionable suggestions for changing inputs to achieve desired outcomes \\
\bottomrule\noalign{}
\end{tabular}
\end{table*}

However, post-hoc explanations of black-box models face fundamental limitations. As \cite{rudin2019} argues, explanations of complex models may not faithfully represent actual model reasoning, potentially providing false comfort. This has driven interest in inherently interpretable models that sacrifice some predictive performance for transparency.

This limitation has catalyzed the development of ``glass-box'' models---architectures designed for interpretability from the ground up rather than requiring post-hoc explanation. Unlike black-box models that necessitate complex explanation techniques, glass-box models make their decision processes transparent by design, a critical advantage for regulated domains like property valuation where every conclusion must be defensible and auditable (\autoref{tab:XAI}).

Generalized Additive Models (GAMs) exemplify this approach, extending traditional linear models by allowing non-linear relationships while maintaining additive separability. Each feature's contribution can be directly visualized and understood. For real estate applications, GAMs might reveal:

\begin{itemize}
\item
  Living area contribution follows a curve with steady increases up to 3,000 sq ft, then diminishing returns
\item
  Age depreciation is steepest in the first 15 years, then moderates
\item
  Bathroom count shows positive but saturating effects after 3 bathrooms
\item
  Lot size premiums vary by density---more valuable in urban areas, less in rural
\end{itemize}

The landscape of modern machine learning models for valuation can be understood through three interpretability paradigms (\autoref{tab:XAI-models}). Traditional black-box models like deep neural networks prioritize predictive accuracy but require complex post-hoc explanation methods. White-box models such as linear regression offer complete transparency but often lack the flexibility to capture complex market dynamics. Glass-box models represent an optimal middle ground, balancing predictive power with inherent interpretability---a crucial requirement for regulatory compliance in property valuation.

\begin{table*}[htbp]
\caption{Comparison of Model Interpretability Paradigms for Property Valuation}
\label{tab:XAI}
\centering
\footnotesize
\begin{tabular}[]{@{}p{0.20\textwidth}p{0.25\textwidth}p{0.25\textwidth}p{0.30\textwidth}@{}}\toprule\noalign{}
Characteristic
 & Black-Box Models
 & White-Box Models
 & Glass-Box Models
 \\
\midrule\noalign{}
Interpretability Level & Low & High & Moderate-to-High (by design) \\
Transparency & Opaque & Fully Transparent & Inherently Transparent \\
Typical Examples & Deep Neural Networks (complex CNNs, RNNs, traditional GNNs) & Linear Regression, Simple Decision Trees & GAMs, EBMs, GNANs \\
Primary Focus & Predictive Accuracy & Interpretability & Balance of Accuracy and Interpretability \\
Explanation Method & Post-hoc (e.g., SHAP, LIME) & Intrinsic (direct model inspection) & Intrinsic (direct model inspection/visualization) \\
Suitability for Regulated Domains & Low (requires significant post-hoc effort) & High (but often lower accuracy) & High (optimal balance for high-stakes decisions) \\
\bottomrule\noalign{}
\end{tabular}
\end{table*}

The evolution of intrinsically interpretable models extends to more complex architectures like Graph Neural Networks (GNNs), increasingly used for modeling spatial relationships between properties in neighborhoods. While traditional GNNs suffer from opacity, Graph Neural Additive Networks (GNANs) by \cite{bechler-speicher2024} offer a promising glass-box alternative. GNANs combine the power of GNNs in capturing complex dependencies within graph-structured data with the interpretability of additive models, providing transparent insights into how neighborhood features influence individual property valuations.

\begin{table*}[htbp]
\caption{Glass-Box Machine Learning Models for Interpretable Property Valuation}
\label{tab:XAI-models}
\centering
\footnotesize
\begin{tabular}[]{@{}p{0.07\textwidth}p{0.17\textwidth}p{0.17\textwidth}p{0.17\textwidth}p{0.17\textwidth}p{0.20\textwidth}@{}}\toprule\noalign{}
Model
 & Core Interpretability Mechanism
 & Key Benefit for Real Estate Appraisal
 & Relevant Context/Application
 & Specific Problems Addressed
 & Alignment with UAD 3.6 Requirements
 \\
\midrule\noalign{}
GAMs & Additive shape functions for each feature; direct visualization & Captures non-linear effects transparently; clear property insights & General valuation; hedonic models; feature impact analysis & Balancing accuracy with interpretability; bias detection & Supports granular data fields; aids narrative explanations \\
EBMs & Extends GAMs with pairwise interactions; intrinsic interpretability & Detailed explanations including feature interactions & Complex valuations; market dynamics understanding & Black-box nature of accurate ML; regulatory compliance & Visualizes interaction effects; enhances verifiability \\
M-GAM & Direct incorporation of missingness indicators with regularization & Handles missing data interpretably; preserves sparsity & Incomplete property records; transparent valuations despite gaps & Bias from imputation; model complexity from missingness & Enables transparent reporting on incomplete data \\
GNANs & Univariate distance and feature shape functions for graph data & Interprets spatial dependencies; neighborhood effects & Modeling location impacts; spatial valuation; geospatial analysis & Overlooking spatial interdependencies; GNN opacity & Supports detailed market influence reporting \\
\bottomrule\noalign{}
\end{tabular}
\end{table*}

These glass-box approaches represent a significant step toward developing AI systems that are not only accurate but also inherently trustworthy and auditable. By moving from opaque black boxes to transparent glass boxes, these models empower appraisers with actionable insights and facilitate compliance with regulatory frameworks like USPAP and UAD 3.6.

\begin{figure}
    \centering
    \includegraphics[width=1\linewidth]{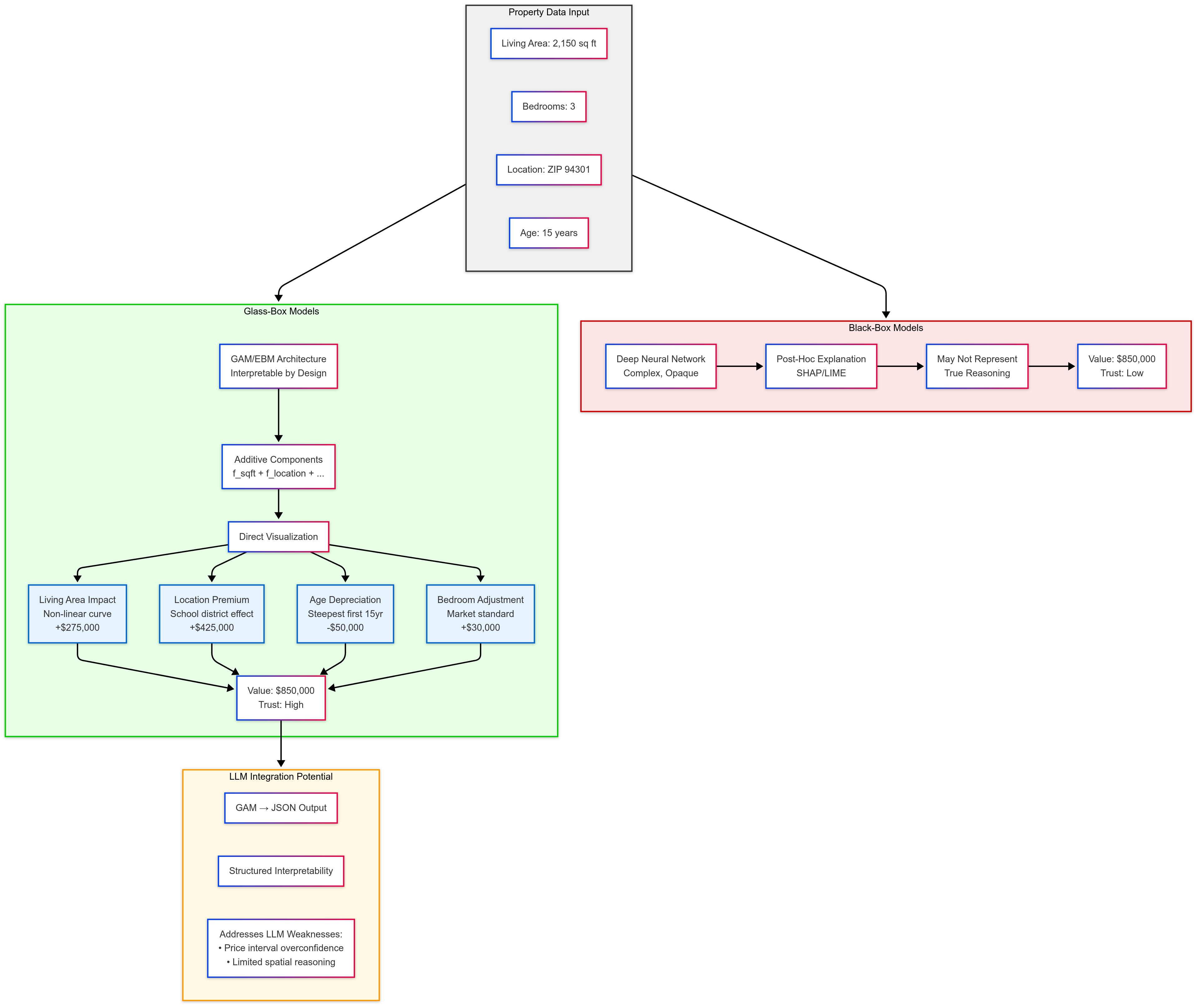}
    \caption{Glass-box models like GAMs and EBMs provide inherent interpretability through additive decomposition of valuation factors, enabling appraisers to directly visualize and validate each component's contribution rather than relying on potentially misleading post-hoc explanations of black-box neural networks. The transparent architecture aligns naturally with traditional appraisal methods where adjustments are explicitly calculated and summed, while the structured outputs can be encoded as JSON for integration with LLMs, addressing their known weaknesses in quantitative reasoning and spatial analysis. This architectural choice embodies the paper's argument that in regulated domains requiring documented reasoning for every conclusion, sacrificing marginal predictive performance for transparency creates systems that are not only accurate but also trustworthy and auditable.}
    \label{fig:glassbox}
\end{figure}

The integration potential between glass-box models and large language models offers promising directions. \citep{lengerichLLMsUnderstandGlassBox2023a} demonstrate how structured outputs from GAMs can be encoded into formats readily understandable by LLMs, such as JSON objects (\autoref{fig:glassbox}). This structured interpretability directly addresses known limitations of LLMs in quantitative domains, particularly their struggles with overconfidence in price intervals and limited spatial reasoning \citep{geerts2025}. A GAM can precisely show the non-linear impact of features like distance to amenities, providing the quantitative grounding that an LLM might otherwise lack. This integration could extend to developing intelligent data scientist agents \citep{bordt2024} as part of agentic AI appraiser frameworks, as demonstrated in other domains \citep{gridach2025, moss2025}.

\subsection{Algorithmic Justice and Historical Redress}

The application of machine learning to real estate valuation occurs against a deeply troubling historical backdrop of systematic discrimination in American housing markets. From explicit redlining policies of the 1930s-1960s that denied mortgages to minority neighborhoods, through subtler forms of steering and discrimination that persist today, real estate data carries the embedded traces of structural racism. As \cite{rothsteinColorLawForgotten2018a} documents comprehensively, these were not merely private prejudices but government-sponsored policies that shaped metropolitan development patterns and wealth accumulation opportunities for generations.

The legacy of these practices persists in contemporary data. Properties in historically redlined areas---predominantly Black and Hispanic neighborhoods---continue to appraise for less than equivalent properties in historically white areas, even after controlling for objective characteristics. This isn't merely correlation but the direct result of decades of disinvestment, where denied mortgages meant deferred maintenance, limited renovation, and reduced buyer competition. Infrastructure investments favored white suburbs while minority neighborhoods faced highway construction and industrial zoning. School funding mechanisms tied to property values created self-reinforcing cycles where lower values meant worse schools, further depressing values.

For AI systems trained on historical data, the risk of perpetuating these patterns is severe. Machine learning models excel at finding patterns in data---if minority neighborhoods have systematically received lower valuations, models will learn to continue this discrimination. The challenge is compounded by proxy variables; even if race is excluded from models, algorithms can infer it through correlated features like zip codes, even linguistic patterns in property descriptions.

The computer science literature on algorithmic fairness offers multiple, often mathematically incompatible, definitions of fairness \citep{narayanan2018}. Demographic parity requires equal outcomes across groups---the same percentage of properties in Black and white neighborhoods should appraise above contract price. Equalized odds demands equal error rates---overvaluation and undervaluation should occur at similar rates across groups. Individual fairness insists similar properties receive similar treatment regardless of neighborhood demographics. In valuation contexts, these definitions conflict: ensuring equal average valuations across neighborhoods (demographic parity) may require treating similar individual properties differently (violating individual fairness).

Resolution requires moving beyond purely statistical approaches to consider causal structures underlying discrimination. \citep{pearlBookWhyNew2020} causal framework provides tools for distinguishing legitimate factors from discriminatory proxies. Neighborhood racial composition should not causally affect property value in a fair market---any observed correlation reflects historical discrimination rather than legitimate value drivers. Counterfactual fairness, as formalized by \cite{kusnerCounterfactualFairness2017}, requires that changing only protected attributes (imagining the same property in a white versus Black neighborhood) should not change outcomes.

Technical approaches to bias mitigation must operate throughout the machine learning pipeline, beginning with pre-processing interventions that modify training data to remove discriminatory patterns. These interventions employ multiple complementary strategies: re-sampling techniques ensure proportional representation across neighborhoods, preventing models from learning that certain areas inherently have lower values. Synthetic data generation takes this further by creating artificial examples of high-value properties in historically undervalued areas, actively teaching models that quality properties can exist anywhere regardless of historical patterns. Feature engineering approaches work to remove or obscure discriminatory proxies, such as replacing discrete neighborhood boundaries with distance gradients that capture location effects without enabling direct racial inference. However, as \citep{dwork2012} demonstrate, simple demographic blindness proves insufficient because algorithms inevitably learn proxies through correlated features---a phenomenon requiring more sophisticated interventions.

The in-processing phase incorporates fairness directly into model training through increasingly sophisticated methods. Adversarial debiasing, as developed by \cite{zhang2018}, exemplifies this approach by training a primary valuation model while simultaneously training an adversary whose goal is to predict protected attributes from the model's internal representations. The primary model must learn to fool this adversary, forcing it to create representations that preserve relevant valuation information while becoming unable to distinguish between properties in different racial neighborhoods. This elegant game-theoretic approach ensures fairness emerges naturally from the training process rather than being imposed externally. Complementing this, fairness-constrained optimization directly incorporates disparate impact measures into the loss functions that guide model training, explicitly trading some predictive accuracy for reduced discrimination---a trade-off that acknowledges the social responsibilities of automated valuation systems.

Post-processing adjustments represent the final opportunity to modify model outputs to achieve fairness metrics, though these approaches require careful implementation to maintain credibility. Threshold optimization techniques find different decision boundaries for different demographic groups to equalize error rates, ensuring that overvaluation and undervaluation occur at similar rates across neighborhoods. Output calibration goes further by ensuring predictions maintain equal accuracy across demographics, not just equal error rates. While conceptually simpler than modifying training procedures, post-processing approaches risk creating obvious disparities that could undermine system credibility if stakeholders perceive that identical properties receive different valuations based solely on neighborhood demographics. The challenge lies in implementing these adjustments subtly enough to achieve fairness goals while maintaining the appearance and reality of objective valuation.

The deeper challenge involves determining what constitutes fairness in valuation contexts. Should algorithms perpetuate existing price differentials that may reflect historical discrimination? Or should they actively counteract such patterns, potentially conflicting with market prediction objectives? Rawlsian principles of justice \citep{rawls1958} might suggest prioritizing the least advantaged---ensuring valuation systems don't further disadvantage historically discriminated communities. Utilitarian approaches might focus on overall market efficiency while constraining discriminatory outcomes.

\section{Uncertainty Quantification and Risk Management in AI-Augmented Valuation}

The transition from traditional point estimates to uncertainty-aware valuations represents a fundamental reconceptualization of property appraisal practice. Traditional valuation reports present single numbers with implied precision that \citep{mallinson2000} argue fundamentally misleads users about inherent market uncertainties and impedes appropriate risk management. This concern extends beyond theoretical accuracy to practical market functioning, as uncertainty in property valuation has emerged as a critical research domain precisely because conventional approaches fail to capture the complex, multifaceted nature of valuation risk \citep{kucharska-stasiakUncertaintyPropertyValuation2014a}. While appraisers have always implicitly acknowledged uncertainty through qualifying statements and professional judgment, AI systems now enable explicit quantification that fundamentally transforms risk assessment throughout the mortgage finance ecosystem.

\subsection{Beyond the Epistemic-Aleatoric Dichotomy}

Machine learning traditionally partitions uncertainty into aleatoric (irreducible randomness) and epistemic (reducible knowledge gaps) components, a framework comprehensively reviewed by \cite{hllermeier2021}. This elegant theoretical distinction has guided uncertainty quantification research for decades, promising to separate what cannot be known from what could be known with more data.

Yet groundbreaking empirical evidence now challenges this foundational assumption. Across 19 different uncertainty estimators and 13 diverse tasks, \citep{smithRethinkingAleatoricEpistemic2025} discovered that aleatoric and epistemic estimates exhibit correlations between 0.8 and 0.999---essentially measuring the same underlying phenomenon regardless of method. Whether using deep ensembles, evidential deep learning, Laplace approximations, or Monte Carlo dropout, the promised decomposition fails to materialize in practice, rendering theoretical distinctions meaningless for real applications.

This empirical collapse aligns with deeper theoretical critiques questioning the conceptual coherence of the framework itself. The aleatoric-epistemic dichotomy suffers from fundamental ambiguity, with epistemic uncertainty alone receiving multiple incompatible mathematical definitions ranging from variance-based to information-theoretic to distance-based measures \citep{smith2024}. Such definitional confusion reflects a deeper problem: attempting to force the rich complexity of uncertainty into two rigid categories creates conceptual overloading that obscures rather than clarifies.

Property valuation exemplifies these theoretical limitations through practical paradoxes. When identical homes on the same street trade at different prices, the variation stems partly from negotiation randomness and buyer preferences---seemingly aleatoric factors. Yet market participants perceive and respond to these uncertainties differently depending on whether they believe the variation reflects unknowable randomness or learnable patterns, with profound implications for investment behavior \citep{walters2023}. This behavioral reality demands uncertainty decomposition that mathematical methods cannot reliably deliver.

\subsection{The Perils of Decoupled Uncertainty Estimation}

Beyond simple correlation, the impossibility of meaningful decomposition stems from fundamental interaction effects between uncertainty types. Epistemic uncertainty about model parameters necessarily creates uncertainty about the true aleatoric distribution---if we don't know the model, we cannot know its inherent randomness. This circular dependency becomes particularly problematic for unique properties or market disruptions where epistemic uncertainty peaks precisely when reliable aleatoric estimates matter most.

\begin{figure}
    \centering
    \includegraphics[width=1\linewidth]{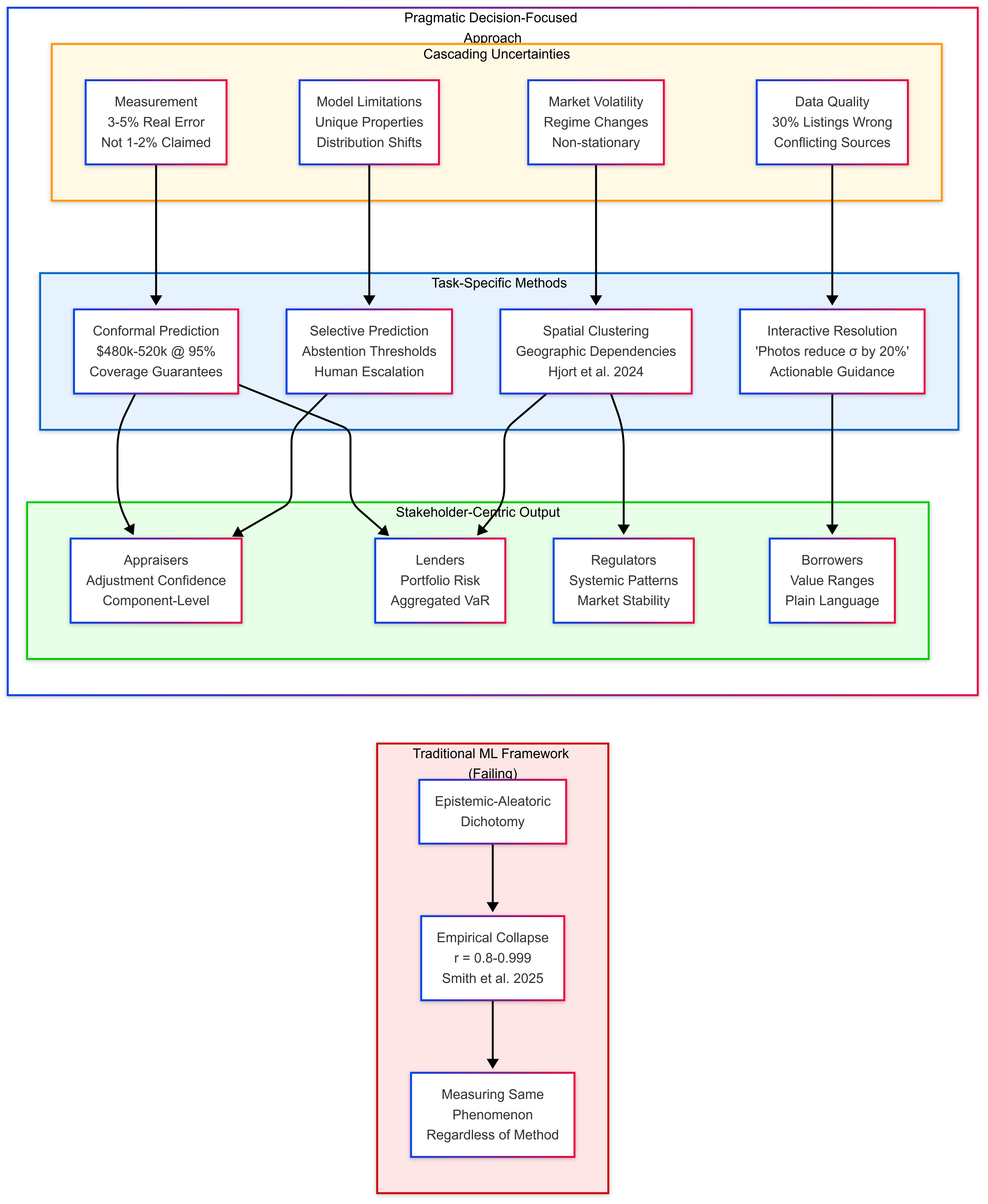}
    \caption{The empirical collapse of the epistemic-aleatoric dichotomy, with correlations exceeding 0.8 across 19 different uncertainty estimators, demands abandoning theoretical purity for pragmatic, decision-focused approaches tailored to specific stakeholder needs. Rather than pursuing impossible decomposition of uncertainty sources that interact multiplicatively in property valuation (measurement error × market volatility × model limitations), effective systems provide actionable intelligence through conformal prediction intervals, spatial clustering, and interactive guidance. This framework operationalizes the paper's argument that in financial markets where valuation errors cascade through leverage into systemic risks, pragmatic uncertainty quantification that acknowledges irreducible ambiguity while quantifying what we can protects against both algorithmic overconfidence and theoretical overreach.}
    \label{fig:uq}
\end{figure}

Property valuation demonstrates these interactions through cascading uncertainties that resist separation (\autoref{fig:uq}). Bayesian neural network approaches to valuation reveal how measurement uncertainty, market volatility, and model limitations interact multiplicatively rather than combining additively \citep{leeRepresentingUncertaintyProperty2020a}. A listing photo of a ``recently renovated kitchen'' embodies simultaneous uncertainties: image quality affects feature detection, semantic ambiguity clouds ``recent'' and ``renovated,'' while market impact remains contested---each source amplifying others in irreducible ways.

Sophisticated deep learning methods promised to untangle these interactions. Deep ensembles capture model uncertainty through prediction disagreement \citep{lakshminarayanan2017}, while Monte Carlo dropout offers computationally efficient Bayesian approximation \citep{gal2016}. Yet even these advanced approaches produce highly correlated uncertainty estimates that fail to achieve meaningful separation, leading \citep{smithRethinkingAleatoricEpistemic2025} to advocate abandoning fixed categories entirely in favor of specialized, task-specific uncertainty quantification.

\subsection{Large Language Models: New Frontiers in Uncertainty}

The integration of large language models into valuation workflows has exposed additional limitations while introducing novel uncertainty challenges. Traditional uncertainty frameworks lose coherence entirely in interactive LLM settings, where popular definitions of aleatoric and epistemic uncertainty directly contradict each other \citep{kirchhof2025}. These authors propose replacing the classical dichotomy with categories better suited to generative AI: underspecification uncertainties arising from incomplete user inputs, interactive uncertainties reducible through dialogue, and output uncertainties expressed through natural language rather than numerical confidence.

LLM uncertainty quantification reveals fundamental challenges distinct from traditional predictive models. Token-level probabilities provide notoriously unreliable uncertainty estimates, with models expressing high confidence while generating factually incorrect statements \citep{yadkoriBelieveNotBelieve2024}. The distinction between semantic uncertainty (whether text conveys intended meaning) and syntactic confidence (word choice probability) requires entirely new quantification approaches that move beyond numerical metrics to linguistic expression \citep{yeBenchmarkingLLMsUncertainty2024}.

Valuation narratives compound these challenges through multiple uncertainty layers that resist traditional categorization. Market analysis claiming ``steady appreciation due to commercial development'' embodies factual uncertainty about appreciation rates, causal uncertainty linking development to values, and linguistic uncertainty around ``steady''---none fitting neatly into classical frameworks. These interconnected uncertainties demand richer representations than binary decomposition allows.

\subsection{Practical Uncertainty Quantification for Valuation}

Given these theoretical impossibilities and practical limitations, effective uncertainty quantification must abandon universal decomposition in favor of decision-focused approaches. Conformal prediction offers a powerful alternative paradigm, constructing prediction sets with guaranteed coverage properties without requiring uncertainty source separation \citep{angelopoulos2021}. For property valuation, this translates to confidence intervals like ``\$480,000-\$520,000 with 95\% coverage'' that remain valid regardless of whether uncertainty stems from market volatility or model limitations.

Spatial dependencies crucial to real estate markets require specialized treatment beyond generic uncertainty methods. By incorporating geographic clustering into conformal prediction frameworks, \citep{hjort2024} achieve valid uncertainty intervals for automated valuation models that respect spatial correlation without attempting philosophical decomposition. Their approach acknowledges that nearby properties share uncertainties through market factors, infrastructure, and neighborhood characteristics---dependencies that matter more than uncertainty taxonomy.

Moving beyond traditional frameworks toward task-specific measures, recent work demonstrates how different stakeholders require fundamentally different uncertainty information \citep{pollestadBetterUncertaintyQuantification2024a}. Appraisers need confidence metrics for specific adjustments, lenders require portfolio risk aggregation, while regulators monitor systemic patterns---none mapping cleanly onto theoretical categories. This stakeholder-centric approach prioritizes actionable intelligence over philosophical purity.

Rather than reporting abstract uncertainty levels, systems should connect uncertainty to specific remediation actions. Indicating that ``valuation uncertainty stems from limited recent comparables---expanding search radius to include the adjacent neighborhood could reduce confidence intervals by approximately \$20,000'' provides actionable guidance without requiring users to interpret epistemic versus aleatoric components. This pragmatic framing serves practitioners better than theoretical decomposition.

\subsection{Implementing Robust Uncertainty Quantification}

The comprehensive review by \cite{wang2025} charts a path forward that embraces uncertainty's complexity rather than forcing artificial simplification. Property valuation demands specialized approaches tailored to specific decision contexts rather than universal frameworks.

Selective prediction mechanisms identify cases where models cannot produce reliable estimates, triggering human review for unique properties or unusual market conditions \citep{geifmanSelectiveClassificationDeep2017a}. This approach acknowledges that some valuations require human expertise that no amount of data or model sophistication can replace, providing systematic frameworks for identifying these cases.

Geographic dependencies in uncertainty propagation demand explicit spatial modeling that moves beyond point-wise estimates. Properties in data-rich neighborhoods warrant tighter confidence intervals than those in areas with sparse transactions, reflecting information availability rather than philosophical uncertainty types \citep{hjort2024a}. Such spatial awareness proves essential for risk assessment across portfolios with geographic concentration.

Interactive uncertainty resolution, inspired by advances in LLM agents, enables systems to identify specific information that would most effectively reduce uncertainty \citep{kirchhof2025a}. Rather than passively reporting confidence levels, systems actively guide data collection: ``Interior photos would reduce valuation uncertainty by approximately 20\%'' or ``Clarifying renovation timing could narrow estimates by \$15,000.'' This active approach transforms uncertainty from limitation to actionable intelligence.

Real-world applications demonstrate the value of pragmatic, decision-focused uncertainty quantification. By developing uncertainty-aware investment strategies that respond to decision-relevant metrics rather than theoretical decomposition, \citep{pollestad2025} show how AI systems can mitigate adverse selection in real estate markets. Their success stems from focusing on what uncertainty means for specific choices rather than its philosophical origins.

The practical wisdom embodied in guides for real estate development under uncertainty emphasizes useful communication over perfect quantification \citep{deneufville2018}. This pragmatic philosophy, informed by mounting evidence against traditional decomposition, acknowledges theoretical limitations while delivering practical value through context-appropriate uncertainty representations.

The convergence of empirical evidence against meaningful aleatoric-epistemic separation, conceptual critiques of the framework's coherence, and practical challenges from LLMs and spatial dependencies demands fundamental reconceptualization. Rather than pursuing elegant but impossible decompositions, property valuation must embrace messy, context-dependent approaches providing actionable intelligence. In financial markets where valuation errors cascade through leverage into systemic risks, this pragmatic uncertainty quantification---acknowledging irreducible ambiguity while quantifying what we can---protects against both algorithmic overconfidence and theoretical overreach. The future lies not in forcing uncertainty into rigid categories but in developing rich, task-specific representations that serve human decisions in all their complexity.

\subsection{Systemic Risk and Financial Stability}

The widespread adoption of AI-based valuation systems raises systemic concerns extending beyond individual accuracy or fairness to market-wide stability. These concerns echo broader debates about algorithmic trading in financial markets \citep{white2013} and automated decision-making in critical infrastructure. For real estate---where valuations affect trillions in mortgage debt and household wealth---systemic effects demand careful consideration.

Model homogenization emerges as perhaps the most insidious systemic risk, developing gradually as market participants converge on similar AI models trained on overlapping data sources. This convergence occurs naturally through multiple mechanisms: vendors compete by promoting their superior accuracy, leading to adoption of similar ``best practices''; regulatory guidance inadvertently standardizes approaches by specifying preferred methods; and open-source frameworks lower barriers to entry but concentrate users on common architectures. The result mirrors the ``herding'' behavior documented in analyst forecasts by \cite{bekaert2000} and credit ratings by \cite{hackbarth2012}. When most market participants use similar models, their errors become correlated rather than independent, potentially manifesting as synchronized overvaluation during booms when all models learn optimistic patterns, or synchronized undervaluation during busts when recent negative data dominates training. The financial crisis of 2008 starkly demonstrated how correlated model errors---in that case, systematic underestimation of mortgage default correlations---can cascade through interconnected financial systems to threaten global stability.

Algorithmic feedback loops introduce dynamic instability absent from traditional human-mediated markets. These loops emerge when AI valuations influence listing prices, which then become training data for future models, creating self-reinforcing cycles that can spiral beyond fundamental values. During market upswings, optimistic AI valuations encourage sellers to list at higher prices; these listings, even if they don't result in sales, become part of the data environment training next-generation models, which learn that prices are rising and generate even more optimistic valuations. This positive feedback creates bubbles where rising predictions become self-fulfilling prophecies divorced from underlying economic fundamentals. More dangerously, negative feedback during downturns could accelerate price declines as models trained on recent distressed sales project continued deterioration, potentially triggering mortgage defaults that create more distressed sales in a vicious cycle. Such dynamics differ qualitatively from human-mediated markets where psychological factors like loss aversion, anchoring on purchase prices, and hope for recovery provide natural dampening that slows both rises and falls.

The speed of adjustment possible with algorithmic valuation fundamentally alters market dynamics in ways we're only beginning to understand. Human appraisers update their mental models gradually, incorporating new information through the lens of years of experience and professional judgment. This measured pace, while sometimes criticized as sluggish, provides stability by preventing overreaction to temporary market disruptions. AI systems, in contrast, can retrain on new data within hours or days, rapidly incorporating recent transactions into their models. During normal market conditions, this responsiveness improves accuracy by quickly reflecting genuine changes in buyer preferences or neighborhood desirability. However, during market transitions or external shocks, rapid adjustment might amplify volatility rather than dampening it. The ``flash crash'' of May 6, 2010, when algorithmic trading caused the Dow Jones to plunge nearly 1,000 points in minutes before recovering, demonstrated how automated systems can create destructive feedback loops faster than human intervention can respond \citep{kirilenko2017}. Similar dynamics in property valuation---where declining values in one neighborhood trigger automated markdowns across entire portfolios within days---could destabilize local markets and precipitate broader regional declines.

Data manipulation vulnerabilities create entirely new attack vectors for market manipulation that didn't exist with human appraisers. Research on adversarial examples in machine learning demonstrates how small, carefully crafted perturbations to input data can cause large, predictable changes in model outputs (\citep{goodfellow2015, zeng2024, li2025}). For valuation systems, strategic actors might exploit these vulnerabilities through coordinated actions: inflating comparable sales through wash trades between conspirators, flooding MLS systems with phantom listings at inflated prices that never result in actual sales, systematically manipulating property descriptions to trigger known model biases, or timing transactions to coincide with model retraining cycles for maximum impact. While human appraisers can certainly be deceived or corrupted, the scale and speed of algorithmic systems amplify manipulation impacts---a successful attack on a widely-used model could simultaneously affect thousands of valuations across entire markets.

Addressing these systemic risks requires coordinated responses that balance innovation with stability. Diversity requirements could mandate heterogeneous models across market participants, with regulatory frameworks specifying minimum variation in model architectures, training data sources, or update frequencies. This approach echoes ecological principles where biodiversity provides resilience against disease and environmental shocks---monocultures may be efficient but prove catastrophically fragile when conditions change. Just as agricultural regulations limit single-crop concentration, financial regulators might limit market share for any single valuation model or require major lenders to use multiple independent systems.

Stress testing must evolve beyond traditional scenarios to evaluate AI system behavior under extreme conditions. Current bank stress tests examine capital adequacy under recession scenarios, but AI systems require additional evaluation: How do models perform during rapid price swings that exceed historical training data? What happens when data quality degrades during crises as transaction volume drops? Can systems handle coordinated adversarial attacks designed to exploit known vulnerabilities? Do feedback loops amplify or dampen external shocks? Regular stress tests, analogous to those conducted for bank capital requirements but adapted for algorithmic systems, could identify vulnerabilities before they manifest in actual crises.

Circuit breakers borrowed from equity markets might limit automated markdown severity during disruptions. If AI valuations for a geographic area decline more than predetermined thresholds---perhaps 5\% in a month or 10\% in a quarter---automatic restrictions could engage: requiring human review for all valuations in affected areas, limiting the weight given to recent distressed transactions, extending time delays between model updates and implementation, or temporarily reverting to more stable historical models. Such mechanisms explicitly trade efficiency for stability, accepting slower adjustment to prevent destructive overshooting that could become self-fulfilling.

Transparency requirements enabling effective market monitoring prove essential for maintaining system stability. Regulators need real-time visibility into model concentration to identify when too many participants use similar approaches, correlation patterns revealing whether models move together or independently, emerging anomalies suggesting manipulation or system malfunction, and performance metrics across different market conditions. This might require standardized reporting protocols covering model characteristics, data sources, update frequencies, uncertainty estimates, and correlation measures. Public disclosure of aggregate statistics, while preserving proprietary details, could enable academic research and independent oversight that identifies risks before they threaten stability.

The goal is not preventing AI adoption but ensuring it enhances rather than threatens stability. Historical parallels suggest this is achievable---electronic trading transformed financial markets while maintaining stability through appropriate safeguards. The key lies in proactive risk management rather than reactive regulation after crises emerge.

\section{Evaluation Methodologies for Professional AI Systems}

The evaluation of artificial intelligence systems for professional applications demands fundamentally different approaches than those developed for general-purpose models. The machine learning community's reliance on standardized benchmarks, while driving progress in research settings, creates misaligned incentives and false confidence when applied to specialized professional domains. This section develops comprehensive evaluation methodologies specifically tailored to AI-augmented property valuation, addressing the unique requirements of accuracy, compliance, fairness, and human-AI collaboration in high-stakes financial decision-making.

\subsection{The Failure of Generic Benchmarks}

The inadequacy of generic evaluation metrics for professional AI applications reflects deeper epistemological issues about knowledge, performance, and domain expertise. Standard natural language processing benchmarks such as GLUE \citep{wang2018}, SuperGLUE \citep{wang2019}, and MMLU \citep{hendrycks2020} test broad linguistic and reasoning capabilities through reading comprehension, logical inference, and general knowledge tasks. While these benchmarks prove valuable for comparing general-purpose models, they fundamentally fail to capture the specialized competencies required for professional practice in domains such as property valuation.

The phenomenon of benchmark saturation compounds these limitations in ways that undermine the entire evaluation enterprise. Contemporary large language models routinely achieve near-ceiling performance on established benchmarks, with GPT-4 scoring 86.4\% on MMLU's 57-task evaluation and specialized models exceeding 90\% on reading comprehension tests. Yet this benchmark success correlates poorly with performance on domain-specific tasks that define professional competence. A model achieving perfect scores on general knowledge questions may still struggle with fundamental appraisal concepts, failing to understand the principle of substitution, miscalculating depreciation schedules, or failing to recognize when adjustments exceed reasonable bounds established by professional practice.

This saturation eliminates the discriminative power that benchmarks ostensibly provide. When all competent models cluster near performance ceilings, meaningful comparison becomes impossible, reducing evaluation to a ritual rather than an informative process. Minor differences in scores often reflect optimization artifacts rather than meaningful capability differences, creating what we might term ``benchmark overfitting'' where development efforts focus on marginal improvements to saturated metrics while fundamental limitations in professional applications remain unaddressed.

The data contamination problem presents even more serious validity concerns that strike at the heart of benchmark reliability. Large language models trained on web-scale corpora inevitably encounter test examples during training, whether in verbatim form, paraphrased versions, or conceptually similar variants. Recent systematic analysis by \cite{zhou2023} found contamination affecting 70-90\% of examples in popular benchmarks, manifesting through verbatim memorization of question-answer pairs, recognition of paraphrased versions, and meta-learning of question patterns and expected responses. Even partial contamination dramatically inflates performance metrics, as models achieve high scores through pattern matching rather than genuine understanding.

For specialized domains like property valuation, generic benchmarks exhibit fundamental misalignment with task requirements that renders them not merely inadequate but potentially misleading. The structure-fluency dichotomy illustrates this misalignment: while benchmarks reward grammatical correctness and linguistic fluency, UAD 3.6 compliance demands specific field completion with standardized enumerations where eloquent prose in wrong fields constitutes complete failure. Similarly, the creativity-consistency tension reveals opposing values, as general language tasks often reward novel, creative responses while professional appraisal requires consistent application of established methodologies where creativity in adjustment calculations violates professional standards. The breadth-depth paradox further emphasizes this disconnect, with benchmarks testing encyclopedic knowledge across diverse topics while appraisal requires deep understanding of specific concepts where market dynamics, depreciation methods, and regulatory requirements matter far more than general trivia. Finally, the context-situatedness gap highlights how benchmark questions typically provide self-contained contexts while real appraisals require integrating information across multiple sources, reconciling conflicts, and understanding local market nuances that cannot be captured in isolated test items.

\subsection{Multi-Tiered Evaluation Framework}

\begin{figure*}
    \centering
    \includegraphics[width=1\linewidth]{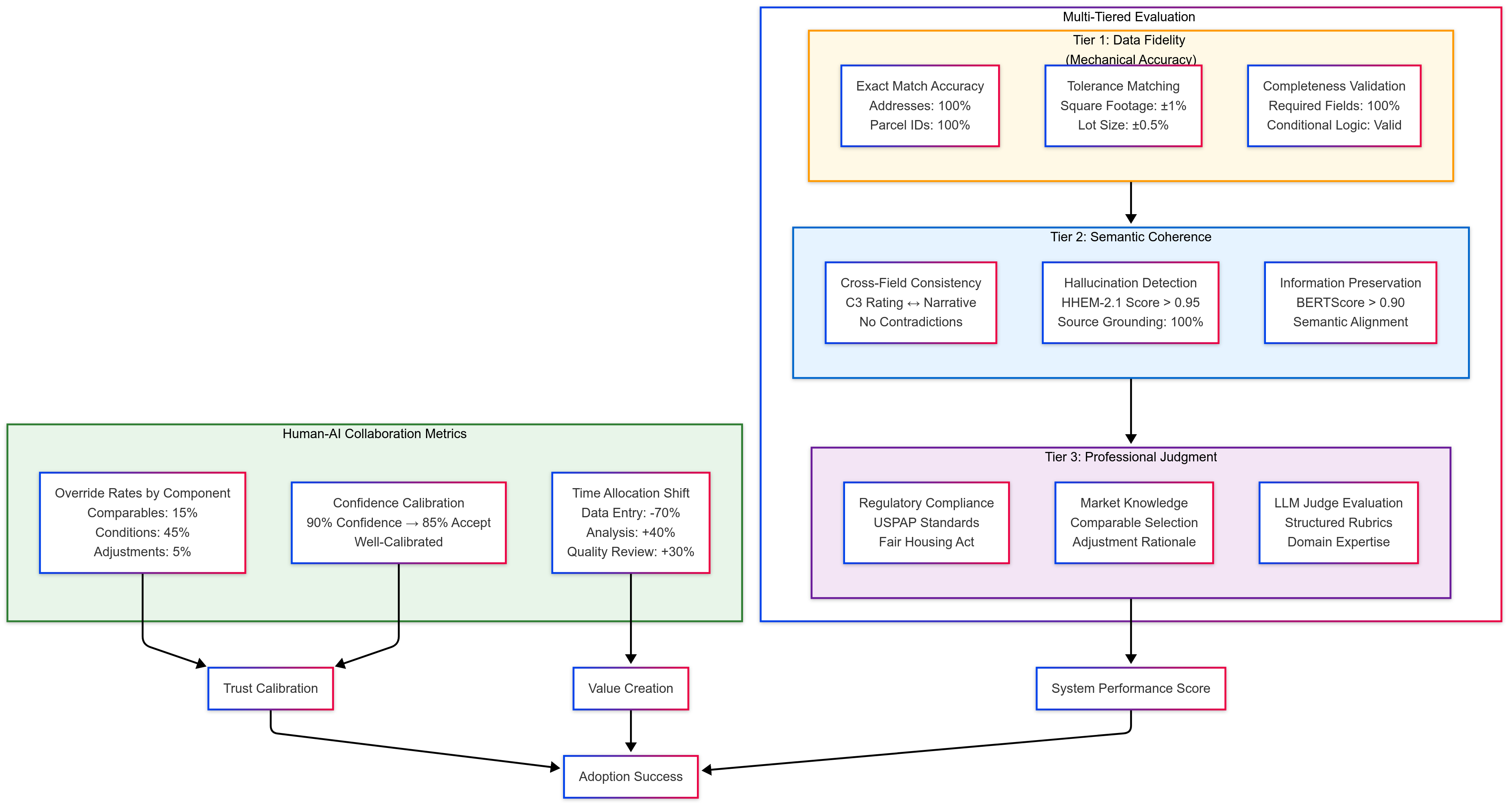}
    \caption{The hierarchical evaluation framework progresses from mechanical accuracy through semantic understanding to professional judgment, with behavioral metrics revealing the quality of human-AI collaboration rather than just technical performance. Unlike generic NLP benchmarks that achieve ceiling effects at 90\%+ accuracy, this domain-specific framework captures the unique requirements of professional valuation including regulatory compliance, market knowledge, and the critical distinction between eloquent prose and correct field completion. The framework operationalizes the paper's key insight that successful evaluation requires not monolithic metrics but multi-dimensional assessment aligned with how professionals actually validate quality, trust systems, and adapt their workflows.}
    \label{fig:evals}
\end{figure*}

Effective evaluation for AI-augmented appraisal systems requires abandoning monolithic metrics in favor of hierarchical assessment frameworks that mirror the complexity of professional practice. We propose a three-tiered evaluation structure that progresses from mechanical accuracy through semantic understanding to professional judgment, with each tier targeting distinct capabilities while building toward comprehensive system validation (\autoref{fig:evals}). This framework draws on \citep{taxonomy2025} of educational objectives, recognizing that professional competence involves not just information recall but increasingly sophisticated cognitive processes.

The foundation of reliable valuation rests on accurate data extraction and field population, forming what we designate as Tier 1: Data Fidelity and Mechanical Accuracy. While conceptually straightforward, this tier proves critical as errors compound through subsequent processing stages in ways that can invalidate entire analyses. A misread address undermines legal validity; incorrect square footage propagates through all calculations, potentially affecting value conclusions by tens of thousands of dollars. This tier evaluates the system's ability to accurately transfer information from diverse sources into structured UAD 3.6 fields, a process that proves more complex than surface appearances suggest.

Evaluation at this foundational level requires sophisticated metrics that recognize the varying nature of different data types. For discrete fields where any deviation constitutes error---property addresses, parcel numbers, critical dates, and enumerated values---exact match accuracy provides the appropriate measure. However, for fields allowing acceptable variation---legal descriptions using different abbreviations while preserving meaning, neighborhood boundaries described differently across sources, or property feature descriptions allowing semantic equivalence---normalized similarity metrics prove more appropriate. Numerical fields require tolerance matching that recognizes measurement precision limits, accepting square footage within \textpm{}1\% for ANSI compliance or lot sizes within \textpm{}0.5\% for survey-grade sources. The completeness dimension adds another layer of complexity, requiring validation that all required fields are populated based on property type, conditional fields are correctly triggered, and internal consistency is maintained across related fields.

Beyond mechanical accuracy lies the realm of semantic understanding, forming Tier 2: Semantic Coherence and Information Preservation. This tier evaluates whether the system preserves meaning through the transformation from source documents to structured reports, a challenge that extends far beyond simple translation. Modern embedding-based metrics such as BERTScore \citep{zhang2019} provide sophisticated semantic comparison by leveraging contextual embeddings to measure similarity between generated and reference text. Unlike surface-level metrics like BLEU that penalize any lexical variation, these approaches recognize semantic equivalence, understanding that ``the property exhibits significant deferred maintenance'' conveys the same meaning as ``the home needs substantial repairs.''

For appraisal applications, semantic coherence evaluation must verify multiple dimensions of consistency and accuracy. Cross-field consistency demands that condition ratings align with narrative descriptions, preventing contradictions where a C3 ``average condition'' rating accompanies descriptions of ``extensive updates'' or ``significant deterioration.'' Market analysis alignment requires that statistical trends support narrative conclusions, ensuring claims of ``rapidly appreciating market'' are backed by data showing actual price increases. Each adjustment must maintain justification coherence, with logical explanations supporting the magnitude of value adjustments through cost basis, paired sales evidence, or market extraction methods. Executive summaries must demonstrate fidelity to detailed findings without introducing new claims or contradicting specifics presented elsewhere in the report.

The critical challenge of hallucination detection---identifying fabricated information unsupported by inputs---becomes particularly acute for generative AI systems in professional contexts. Recent metrics like HHEM-2.1 \citep{vectara2024} score factual consistency between source documents and generated text, but appraisal-specific hallucinations require specialized detection approaches. These may manifest as invented amenities not observed in property documentation, fabricated comparable sales that don't exist in MLS records, mischaracterized property conditions contradicted by inspection evidence, or false historical narratives about property or neighborhood development unsupported by factual records.

The highest evaluation tier, Tier 3: Professional Judgment and Regulatory Compliance, addresses abstract qualities requiring professional expertise that cannot be reduced to mechanical rules. This tier must evaluate regulatory compliance with USPAP and UAD requirements, assess appropriate tone and objectivity, verify logical reasoning in value conclusions, and provide holistic quality assessment. The complexity of these requirements necessitates sophisticated evaluation approaches that can capture nuanced professional judgment while maintaining consistency and reliability.

We propose employing carefully calibrated Large Language Model (LLM) judges for this tier, leveraging their sophisticated reasoning capabilities while addressing known biases and limitations through structured evaluation protocols. The key lies in decomposing professional judgment into specific, measurable dimensions that can be consistently assessed. Fair Housing compliance evaluation must screen for discriminatory language or implications, identifying not just explicit references to protected characteristics but subtle biases in neighborhood descriptions and euphemisms potentially masking discrimination. USPAP adherence assessment verifies sufficient support for value conclusions, required disclosures and certifications, scope of work appropriateness, and competency claims with reference to specific standards and rules. Professional objectivity evaluation examines neutrality of language, distinguishing between advocacy and analysis while assessing balance in presenting positive and negative property attributes. Logical coherence verification ensures conclusions follow from presented data, mathematical accuracy in adjustments, sound reconciliation logic across approaches, and absence of circular reasoning. Market knowledge assessment evaluates understanding of local dynamics, appropriateness of comparable selection, alignment of adjustment magnitudes with market evidence, and awareness of current trends.

\subsection{Human-AI Collaboration Metrics}

Technical performance metrics alone fail to capture the complex dynamics of human-AI collaboration that ultimately determine real-world effectiveness. Drawing on established frameworks from human-computer interaction \citep{wang2019}, computer-supported cooperative work, and collaborative intelligence \citep{wilson2018}, we identify behavioral metrics that reveal the quality and effectiveness of human-AI partnership in professional valuation contexts.

Override rates provide direct behavioral measures of system utility and trust, serving as revealed preferences that complement stated satisfaction measures. Granular analysis of override patterns across different valuation components reveals nuanced insights into system performance and user confidence. When appraisers override comparable selection at a 15\% rate, this suggests generally good performance with identifiable improvement areas, particularly in unique property types such as historic homes or mixed-use properties where standard similarity metrics fail to capture relevant comparability factors. Higher override rates for condition ratings, perhaps reaching 45\%, indicate significant trust issues that investigation might reveal stem from systematic biases, such as AI consistently rating properties one condition level too high due to training on listing photos that emphasize positive features. Conversely, low override rates of 5\% for adjustment calculations demonstrate excellent performance, with overrides primarily occurring for unusual circumstances requiring human judgment such as seller concessions or personal property inclusions.

Temporal analysis of override patterns provides additional insights into system learning and adaptation. Decreasing rates over time indicate successful incorporation of feedback and algorithmic improvement, while sudden spikes suggest data distribution shifts from market disruptions or new property types requiring attention. Plateauing rates reveal fundamental capability limits where human expertise remains essential, helping organizations understand the boundaries of effective automation.

The calibration between AI-expressed uncertainty and human acceptance represents another critical dimension of collaboration effectiveness. Ideal calibration manifests as a diagonal relationship where system confidence directly correlates with user acceptance rates. Miscalibration patterns reveal system limitations that undermine trust and efficiency. Overconfident systems that express 90\% confidence but achieve only 60\% acceptance erode trust rapidly as users learn predictions are unreliable. Underconfident systems expressing 50\% confidence while achieving 80\% acceptance waste human time reviewing routine cases that could be processed automatically. Domain-specific miscalibration, where confidence proves accurate for routine properties but poor for complex cases, suggests the need for context-aware uncertainty quantification.

Time allocation analysis examines how AI assistance reshapes professional work patterns, revealing whether technology merely accelerates existing workflows or enables fundamental transformation. Effective augmentation should demonstrate substantial reduction in routine data gathering and entry tasks, moderate reduction in comparable search and analysis time, minimal change or potential increase in complex property analysis time, and expanded time for quality review and professional development. If AI merely speeds existing workflows without enabling higher-value activities, the transformative potential remains unrealized. The goal extends beyond efficiency to professional enhancement, enabling appraisers to focus on complex judgments, market analysis, and quality assurance that leverage their expertise.

\subsection{Continuous Learning and Meta-Evaluation}

Static evaluation fails to capture the adaptive nature of modern AI systems that improve through deployment and use. Professional deployment requires continuous learning from expert feedback, necessitating evaluation frameworks that assess not just current performance but learning capability and trajectory. This dynamic perspective recognizes that system quality emerges through iterative refinement rather than being fixed at deployment.

Learning occurs across multiple levels with different temporal scales and impacts. At the instance level, individual corrections immediately improve current outputs, such as when an appraiser adjusts a value with specific explanation, enabling real-time system adaptation. Pattern-level learning aggregates feedback to reveal systematic biases, such as consistent adjustment modifications in particular market areas that indicate evolving local preferences. Strategic-level learning identifies long-term patterns that inform architectural changes, where consistent overrides for certain property types might trigger development of specialized models or persistent uncertainty in specific scenarios could drive targeted data collection efforts.

Effective learning requires rich feedback that extends beyond binary acceptance or rejection. Structured corrections with explanations provide learning signals about both what and why changes are needed. Confidence calibration feedback helps systems better express uncertainty even when producing correct values. Contextual exceptions teach systems about local market variations while maintaining general principles. Stylistic preferences help align output with professional communication norms without compromising accuracy.

Most critically, evaluation frameworks themselves must evolve through professional input, creating a co-evolutionary dynamic between AI systems and quality metrics. Appraisers identify quality dimensions that automated metrics miss, such as understanding positive connotations of ``original charm'' in historic districts, recognizing cultural boundaries' importance for property values, or awareness of pending zoning changes known locally but not yet formally documented. Through collaborative refinement, new metrics emerge that capture local market authenticity through appropriate terminology use, defensive documentation depth that anticipates revision requests, professional credibility in maintaining expertise tone without overconfidence, and regulatory prescience in adapting to emerging requirements before formal codification.

This co-evolution of AI systems and evaluation frameworks represents genuine human-AI partnership extending beyond task performance to the definition of excellence itself. The evaluation system learns what matters to professionals while professionals learn to articulate tacit quality judgments in measurable terms, creating a virtuous cycle of improvement that benefits both human experts and AI systems.

\section{Implementation Dynamics and Market Evolution}

The transformation from theoretical framework to market reality involves complex organizational, economic, and social dynamics that determine adoption patterns and ultimate impact. This section analyzes implementation challenges through established lenses of technology diffusion \citep{rogersDiffusionInnovations5th2003a}, organizational change \citep{kotterLeadingChange1996a}, and labor market evolution \citep{acemoglu2018}. We examine how different stakeholders must adapt, the economic forces shaping adoption, and the likely patterns of market evolution within the broader context of professional service transformation.

\subsection{Stakeholder Adaptation and Professional Transformation}

The implementation of AI-augmented valuation systems requires fundamental reorganization across the real estate finance ecosystem. Each stakeholder group faces unique challenges and opportunities that demand strategic adaptation rather than passive acceptance of technological change. This transformation exemplifies the broader phenomenon of professional service evolution in the digital age, where technology not merely automates existing processes but fundamentally reconceptualizes professional roles and value creation.

For appraisers, AI augmentation represents the most profound transformation since the profession's formalization in the 1980s following the savings and loan crisis. The shift from data collection to analysis fundamentally alters the profession's value proposition and required competencies, following patterns observed in other professions undergoing technological disruption. As radiologists evolved from film reading to diagnostic integration, accountants from bookkeeping to strategic advisory, and lawyers from document review to complex negotiation, appraisers must navigate a similar transition that preserves professional expertise while embracing technological capabilities.

The task-based framework of technological change developed by \href{https://doi.org/10.1016/S0169-7218(11)02410-5}{Acemoglu and Autor (2011)} provides crucial insight into this evolution. Appraisal work decomposes into distinct task categories with markedly different automation potential, creating a nuanced landscape of human-AI collaboration rather than simple replacement. Routine cognitive tasks such as physical measurement, data entry, comparable searches, and basic calculations face immediate automation pressure. These tasks, while time-consuming and essential to the valuation process, require limited professional judgment and offer minimal opportunity for value differentiation among practitioners.

In contrast, non-routine analytical tasks remain fundamentally human-centric while benefiting from AI augmentation. Complex property analysis for unique features, market trend interpretation requiring local knowledge, reconciliation across valuation approaches, and quality assurance of automated outputs demand the sophisticated judgment that defines professional expertise. These tasks benefit from AI support through enhanced data access and computational power while requiring human insight for final determinations. Non-routine interpersonal tasks resist automation entirely, including property owner interviews, legal testimony, client consultation, and professional networking. These human-centric activities may paradoxically become more valuable as routine tasks face automation, concentrating human effort on irreducibly social aspects of professional practice.

The implications for human capital investment prove profound and far-reaching. Traditional appraisal education emphasized measurement techniques, form completion, and regulatory compliance---skills increasingly obsolete in an AI-augmented environment. Future-oriented professionals must develop complementary capabilities that leverage rather than compete with technological tools. Technology fluency extends beyond software operation to conceptual understanding of AI capabilities and limitations, recognition of appropriate trust calibration, interpretation of uncertainty estimates, and effective direction of AI assistants. Enhanced analytical capabilities become primary value drivers as routine calculations face automation, with professional differentiation emerging through sophisticated market analysis, nuanced adjustments for unique features, narrative skills explaining complex valuations, and pattern recognition identifying market shifts before algorithmic detection.

The profession faces likely bifurcation between high-skill specialists and commoditized generalists, creating distinct career trajectories with different economic prospects. Specialization opportunities emerge in niches where human expertise remains irreplaceable: historic properties requiring preservation knowledge, complex commercial properties with multiple revenue streams, litigation support demanding courtroom credibility, portfolio analysis for institutional clients, and market transition analysis during disruptions. Geographic arbitrage becomes possible as analytical work decouples from physical presence, though this risks degrading local market knowledge essential for quality control.

Yet critical research challenges remain in understanding optimal human-AI collaboration patterns. How professionals calibrate trust in AI systems, what interfaces best support human oversight, and how systems can learn from implicit expert feedback represent open questions requiring interdisciplinary investigation. The development of effective collaboration frameworks will determine whether AI augments or frustrates professional practice.

Appraisal Management Companies face particularly acute transformation pressure as their traditional value proposition becomes increasingly obsolete. The fundamental business model evolution from labor intermediaries to technology platforms represents a stark choice between adaptation and extinction. Technology infrastructure becomes the core offering rather than ancillary service, requiring investment in unified dashboards integrating multiple AI providers, standardized APIs enabling seamless data flow, automated quality control systems, and real-time analytics providing market intelligence.

The vendor ecosystem orchestration function replaces simple appraiser panel management, with modern AMCs coordinating complex networks of specialized providers. This includes data providers offering MLS access, public records, and imagery services; AI service providers delivering various specialized models; quality assurance systems combining automated and human review; compliance tools for fair lending testing and regulatory verification; and integration with client systems including lender platforms and GSE infrastructure. Value-added analytics differentiate commodity services by providing portfolio risk assessment, market trend forecasting, anomaly detection for fraud prevention, performance benchmarking, and custom reporting for enterprise clients.

Financial institutions stand to capture the most direct benefits from AI-augmented appraisal but face substantial organizational challenges in realizing this potential. Legacy systems designed for document-based workflows require fundamental modernization to support structured data processing, confidence interval handling, real-time API integration, and exception-based routing. The technical challenges pale compared to process reengineering requirements that challenge established workflows and departmental boundaries.

Organizational restructuring becomes inevitable as traditional distinctions between processing, underwriting, and quality control blur when AI systems perform initial assessments. New organizational models emerge featuring cross-functional teams managing end-to-end workflows, centers of excellence for AI governance, dedicated roles for model validation and monitoring, and hybrid positions combining technical and domain expertise. Risk model enhancement leverages richer property data for sophisticated assessment beyond traditional variables, incorporating uncertainty-aware pricing, dynamic collateral assessment, behavioral pattern recognition, and portfolio correlation analysis.

The adoption dynamics in regulated industries like mortgage finance require careful study to understand how regulatory frameworks evolve with technology, what determines successful versus failed implementations, and what interventions accelerate beneficial adoption. These questions extend beyond technical implementation to fundamental organizational change management challenges.

\subsection{Economic Analysis of Transformation}

The economics of AI implementation in valuation extend far beyond simple cost-benefit calculations to encompass network effects, learning curves, and competitive dynamics that fundamentally shape adoption patterns and market structure. Understanding these economic forces proves essential for stakeholders navigating the transformation and policymakers seeking to guide beneficial outcomes.

Surface-level analysis focusing solely on labor substitution dramatically underestimates transformation value, missing the complex web of benefits that emerge through systematic change. Comprehensive return on investment calculations must account for multiple benefit streams across different time horizons, recognizing that value creation occurs through system-wide optimization rather than isolated efficiency gains. Direct operational benefits provide immediate returns through labor productivity gains for routine residential properties, cycle time reduction improving customer satisfaction and enabling volume flexibility, revision rate decreases through automated compliance checking, and elimination of calculation and transcription errors.

Hidden value streams emerge through operational transformation as organizations discover previously impossible capabilities. Supervisory efficiency improves through exception-based management that focuses human attention on high-value decisions. Expert knowledge becomes preserved and democratized as heuristics are encoded in systems accessible to all users. Capacity flexibility enables handling volume spikes without proportional staffing increases, while quality consistency reduces variance across appraisers and geographies. These indirect benefits often exceed direct cost savings but prove harder to quantify in traditional ROI calculations.

Risk mitigation value frequently exceeds operational savings, though its probabilistic nature complicates measurement. GSE representation and warranty relief can be worth 5-10 basis points on conforming loans, translating to substantial value for large originators. Reduced buyback risk from consistent, well-documented valuations prevents costly loan repurchases. Fair lending compliance helps avoid violations carrying both financial penalties and reputational damage. Fraud prevention through pattern detection and anomaly identification protects against losses that can devastate profitability.

Strategic analytics capabilities create competitive advantages that compound over time. Granular portfolio risk assessment enables optimized capital allocation, while market intelligence from aggregated valuation data informs expansion decisions. Product development benefits from detailed property characteristic analysis, and regulatory capital optimization becomes possible through better risk measurement. These strategic benefits create sustainable competitive advantages that justify transformation investments even when operational savings alone might not.

The comprehensive ROI calculation reveals total returns of 300-500\% over 3-5 years for well-executed implementations, but realizing these returns requires navigating the productivity J-curve phenomenon identified by \cite{brynjolfssonProductivityJCurveHow2021}. Organizations must sustain investment through initial productivity declines as systems are implemented and workers retrained, requiring patient capital and committed leadership.

Technology-driven unbundling fundamentally reshapes market structure as the integrated appraisal model fragments into specialized components with different economic characteristics. Commoditized data collection faces severe price pressure as barriers to entry collapse, trending toward gig economy models with wages approaching minimum viable levels. Analytical services command premium pricing as professionals demonstrating superior capabilities capture increasing value through volume leverage. Technology platforms exhibit winner-take-all dynamics through high fixed costs and network effects, while boutique specialists find protected niches through differentiation and relationship capital.

First-mover advantages compound through multiple mechanisms creating competitive moats difficult for later entrants to overcome. Data network effects improve model quality with scale, as each processed appraisal enhances algorithms and creates quality gaps competitors struggle to close. Talent concentration accelerates as skilled professionals gravitate toward technology leaders, creating virtuous cycles of improvement. Ecosystem lock-in emerges as workflows integrate deeply, with switching costs compounding through retraining requirements and process disruption. Early movers risk regulatory capture by influencing standards in ways that create subtle but persistent advantages.

The analysis of market structure evolution under AI transformation reveals critical questions requiring ongoing research. Will winner-take-all dynamics dominate or will specialized niches persist? How do network effects interact with regulatory requirements to shape competitive landscapes? What governance mechanisms can prevent harmful concentration while preserving innovation incentives? These questions demand continuous monitoring as markets evolve.

\subsection{Labor Market Implications}

The labor market implications of AI-augmented valuation extend beyond individual job displacement to fundamental restructuring of career paths, skill requirements, and professional identity. These changes exemplify broader patterns identified in the automation and employment literature while exhibiting profession-specific nuances that shape individual and collective responses.

The appraisal profession exhibits classic job polarization patterns identified by \cite{autor2003}, with middle-skill routine jobs facing automation pressure while both high-skill analytical and low-skill manual tasks remain human-centric. This creates an hourglass-shaped occupation structure that fundamentally alters career trajectories and economic prospects. The high-skill segment comprising valuation consultants, review appraisers, and expert witnesses commands premium compensation for complex analytical work requiring deep expertise, sophisticated judgment, and communication skills. Career paths in this segment emphasize specialization and reputation building, with compensation potentially exceeding current averages as volume leverage multiplies earning potential.

Conversely, the low-skill segment of field data collectors and photography assistants performs standardized capture tasks requiring basic technology skills but limited valuation expertise. Compensation faces downward pressure from expanded labor supply and gig economy dynamics, with career advancement requiring transition to analytical roles through education and experience. The hollowing middle of traditional residential appraisers performing full-service valuations faces the greatest pressure, neither specialized enough for premium segments nor willing to accept commoditized wages. This group must adapt or exit, facing not just economic pressure but the psychological challenge of professional identity loss.

Geographic disparities intensify through technology as AI enables separation of locally-bound data collection from location-independent analysis. This creates opportunities for regional arbitrage where skilled analysts in low-cost areas review properties in high-cost markets, though this risks degrading local knowledge essential for quality valuations. Urban concentration of complex properties and specialized services may deplete expertise in smaller markets, while digital divides create valuation deserts in areas with poor connectivity or technology adoption, potentially requiring regulatory intervention.

The aging appraiser demographic complicates adaptation significantly. With average ages exceeding 50, many practitioners face difficult choices between investing in new skills versus pursuing early retirement. Traditional apprenticeship models break down when routine tasks that taught fundamentals face automation, requiring fundamental reimagining of professional education and development. Successful transition requires comprehensive support including educational transformation shifting from measurement techniques to analytical methods, transition assistance through retraining programs and income support, and creation of alternative pathways that preserve valuable expertise while acknowledging changed economics.

Critical questions emerge about how professional cultures adapt to AI augmentation and what interventions most effectively support beneficial transitions. Research from other industries suggests that successful workforce transformation requires early communication, comprehensive support programs, multiple adaptation pathways, and gradual implementation timelines. Organizations must balance efficiency gains with social responsibility, recognizing that abrupt displacement creates costs extending beyond individual hardship to societal disruption.

The temporal modeling challenges for dynamic labor markets add another layer of complexity. Real estate markets exhibit seasonal patterns, economic cycles, and structural breaks that affect employment needs. Research into how AI systems can learn transferable patterns across markets while adapting quickly to regime changes becomes essential for workforce planning. The ability to distinguish signal from noise in volatile periods while maintaining stability determines whether technology creates sustainable employment or destructive churn.

\subsection{Emerging Challenges and Evolution Pathways}

As implementation accelerates, critical challenges emerge that will shape the trajectory of AI-augmented valuation and determine whether transformation enhances or undermines market functioning. These challenges span technical, organizational, and societal dimensions, requiring coordinated responses from multiple stakeholders.

End-to-end uncertainty quantification remains a fundamental technical challenge with profound practical implications. Current methods handle uncertainty at individual stages---measurement accuracy, model confidence, market volatility---but struggle with propagation through complex analytical pipelines. Without comprehensive uncertainty modeling, systems risk false precision that misleads users and regulators. Key research needs include modeling dependencies between uncertainty sources, developing computationally efficient propagation methods, creating interpretable representations for diverse stakeholders, and handling non-stationary uncertainty as markets evolve. Progress in these areas will determine whether AI systems can provide the nuanced risk assessment that sophisticated financial decisions require.

The fundamental limitation of correlation versus causation in AI systems proves particularly problematic for policy-relevant applications. While current models excel at pattern identification, they struggle with causal understanding critical for valuation analysis and market intervention. A system might accurately predict that properties near new transit stations appreciate, but without causal understanding, it cannot distinguish whether proximity causes appreciation or whether transit planners strategically locate stations in appreciating areas. Advances require learning causal structures from observational data, handling confounders in non-experimental settings, enabling counterfactual reasoning for scenario analysis, and maintaining causal validity under distribution shifts that characterize evolving markets.

Systemic risks from widespread AI adoption demand proactive management to prevent market instability. Model homogenization represents a primary concern---if market participants converge on similar models trained on overlapping data, correlated errors could amplify rather than cancel. This herding behavior could manifest as synchronized overvaluation during booms or undervaluation during busts, potentially exceeding the systemic risks that human appraisers with diverse perspectives naturally mitigate. Addressing these risks requires diversity requirements mandating heterogeneous models, regular stress testing under extreme scenarios, circuit breakers limiting automated markdown severity, and transparency enabling market monitoring.

The speed of AI-enabled adjustment fundamentally changes market dynamics in ways requiring careful study. While human appraisers update mental models gradually, AI systems can retrain rapidly on new data. This responsiveness improves accuracy during normal conditions but might amplify volatility during transitions. Research into optimal update frequencies, dampening mechanisms, and human oversight triggers becomes essential for maintaining market stability while capturing efficiency benefits.

Adversarial robustness presents novel vulnerabilities as AI systems create new attack vectors for market manipulation. Research on adversarial examples demonstrates how small input perturbations can cause large output changes in neural networks. For valuation systems, strategic actors might inflate comparable sales through coordinated transactions, flood databases with misleading listings, or exploit known model biases. Developing game-theoretic defenses against such manipulation while maintaining system usability represents an ongoing challenge requiring collaboration between security researchers and domain experts.

The evolution of human-AI collaboration patterns in professional contexts remains poorly understood despite its critical importance. As systems become more capable, the risk of automation bias---over-reliance on AI recommendations---increases. Conversely, excessive skepticism wastes efficiency gains. Research into optimal trust calibration, interface design for effective oversight, and organizational structures supporting collaboration will determine whether AI augments or replaces professional judgment. The development of ``mechanistic interpretability''---understanding not just what models predict but how they reason---offers promising directions for building appropriate professional trust.

International differences in regulatory approaches, market structures, and cultural attitudes toward AI create additional complexity for global organizations. While some jurisdictions embrace technological innovation, others maintain restrictive frameworks prioritizing human oversight. These differences affect competitive dynamics, creating arbitrage opportunities but also compliance challenges. Understanding how to develop AI systems that adapt to local requirements while maintaining global consistency represents an ongoing challenge for multinational participants.

As these challenges illustrate, the transformation of property valuation through AI augmentation is not a deterministic process but one shaped by ongoing choices of technologists, practitioners, regulators, and society. The research frontiers identified throughout this analysis---from technical advances in uncertainty quantification to organizational innovations in human-AI collaboration---will determine whether this transformation enhances professional practice and market efficiency or creates new forms of systemic risk and inequality. Progress demands not just technological sophistication but thoughtful integration of human expertise, regulatory frameworks, and market mechanisms, working together to shape beneficial outcomes from inevitable change.

\section{Implementation Dynamics and Market Evolution}

The transformation from theoretical framework to market reality involves complex organizational, economic, and social dynamics that determine adoption patterns and ultimate impact. This section analyzes implementation challenges through established lenses of technology diffusion \citep{rogersDiffusionInnovations5th2003a}, organizational change \citep{kotterLeadingChange1996a}, and labor market evolution \citep{acemoglu2018}. We examine how different stakeholders must adapt, the economic forces shaping adoption, and the likely patterns of market evolution within the broader context of professional service transformation.

\subsection{Stakeholder Adaptation and Professional Transformation}

The implementation of AI-augmented valuation systems requires fundamental reorganization across the real estate finance ecosystem. Each stakeholder group faces unique challenges and opportunities that demand strategic adaptation rather than passive acceptance of technological change. This transformation exemplifies the broader phenomenon of professional service evolution in the digital age, where technology not merely automates existing processes but fundamentally reconceptualizes professional roles and value creation.

For appraisers, AI augmentation represents the most profound transformation since the profession's formalization in the 1980s following the savings and loan crisis. The shift from data collection to analysis fundamentally alters the profession's value proposition and required competencies, following patterns observed in other professions undergoing technological disruption. As radiologists evolved from film reading to diagnostic integration, accountants from bookkeeping to strategic advisory, and lawyers from document review to complex negotiation, appraisers must navigate a similar transition that preserves professional expertise while embracing technological capabilities.

The task-based framework of technological change developed by \href{https://doi.org/10.1016/S0169-7218(11)02410-5}{Acemoglu and Autor (2011)} provides crucial insight into this evolution. Appraisal work decomposes into distinct task categories with markedly different automation potential, creating a nuanced landscape of human-AI collaboration rather than simple replacement. Routine cognitive tasks such as physical measurement, data entry, comparable searches, and basic calculations face immediate automation pressure. These tasks, while time-consuming and essential to the valuation process, require limited professional judgment and offer minimal opportunity for value differentiation among practitioners.

In contrast, non-routine analytical tasks remain fundamentally human-centric while benefiting from AI augmentation. Complex property analysis for unique features, market trend interpretation requiring local knowledge, reconciliation across valuation approaches, and quality assurance of automated outputs demand the sophisticated judgment that defines professional expertise. These tasks benefit from AI support through enhanced data access and computational power while requiring human insight for final determinations. Non-routine interpersonal tasks resist automation entirely, including property owner interviews, legal testimony, client consultation, and professional networking. These human-centric activities may paradoxically become more valuable as routine tasks face automation, concentrating human effort on irreducibly social aspects of professional practice.

The implications for human capital investment prove profound and far-reaching. Traditional appraisal education emphasized measurement techniques, form completion, and regulatory compliance---skills increasingly obsolete in an AI-augmented environment. Future-oriented professionals must develop complementary capabilities that leverage rather than compete with technological tools. Technology fluency extends beyond software operation to conceptual understanding of AI capabilities and limitations, recognition of appropriate trust calibration, interpretation of uncertainty estimates, and effective direction of AI assistants. Enhanced analytical capabilities become primary value drivers as routine calculations face automation, with professional differentiation emerging through sophisticated market analysis, nuanced adjustments for unique features, narrative skills explaining complex valuations, and pattern recognition identifying market shifts before algorithmic detection.

The profession faces likely bifurcation between high-skill specialists and commoditized generalists, creating distinct career trajectories with different economic prospects. Specialization opportunities emerge in niches where human expertise remains irreplaceable: historic properties requiring preservation knowledge, complex commercial properties with multiple revenue streams, litigation support demanding courtroom credibility, portfolio analysis for institutional clients, and market transition analysis during disruptions. Geographic arbitrage becomes possible as analytical work decouples from physical presence, though this risks degrading local market knowledge essential for quality control.

Yet critical research challenges remain in understanding optimal human-AI collaboration patterns. How professionals calibrate trust in AI systems, what interfaces best support human oversight, and how systems can learn from implicit expert feedback represent open questions requiring interdisciplinary investigation \citep{amershi2019, seeber2020}. The development of effective collaboration frameworks will determine whether AI augments or frustrates professional practice.

Appraisal Management Companies face particularly acute transformation pressure as their traditional value proposition becomes increasingly obsolete. The fundamental business model evolution from labor intermediaries to technology platforms represents a stark choice between adaptation and extinction. Technology infrastructure becomes the core offering rather than ancillary service, requiring investment in unified dashboards integrating multiple AI providers, standardized APIs enabling seamless data flow, automated quality control systems, and real-time analytics providing market intelligence.

The vendor ecosystem orchestration function replaces simple appraiser panel management, with modern AMCs coordinating complex networks of specialized providers. This includes data providers offering MLS access, public records, and imagery services; AI service providers delivering various specialized models; quality assurance systems combining automated and human review; compliance tools for fair lending testing and regulatory verification; and integration with client systems including lender platforms and GSE infrastructure. Value-added analytics differentiate commodity services by providing portfolio risk assessment, market trend forecasting, anomaly detection for fraud prevention, performance benchmarking, and custom reporting for enterprise clients.

Financial institutions stand to capture the most direct benefits from AI-augmented appraisal but face substantial organizational challenges in realizing this potential. Legacy systems designed for document-based workflows require fundamental modernization to support structured data processing, confidence interval handling, real-time API integration, and exception-based routing. The technical challenges pale compared to process reengineering requirements that challenge established workflows and departmental boundaries.

Organizational restructuring becomes inevitable as traditional distinctions between processing, underwriting, and quality control blur when AI systems perform initial assessments. New organizational models emerge featuring cross-functional teams managing end-to-end workflows, centers of excellence for AI governance, dedicated roles for model validation and monitoring, and hybrid positions combining technical and domain expertise. Risk model enhancement leverages richer property data for sophisticated assessment beyond traditional variables, incorporating uncertainty-aware pricing, dynamic collateral assessment, behavioral pattern recognition, and portfolio correlation analysis.

The adoption dynamics in regulated industries like mortgage finance require careful study to understand how regulatory frameworks evolve with technology, what determines successful versus failed implementations, and what interventions accelerate beneficial adoption. These questions extend beyond technical implementation to fundamental organizational change management challenges.

\subsection{Economic Analysis of Transformation}

The economics of AI implementation in valuation extend far beyond simple cost-benefit calculations to encompass network effects, learning curves, and competitive dynamics that fundamentally shape adoption patterns and market structure. Understanding these economic forces proves essential for stakeholders navigating the transformation and policymakers seeking to guide beneficial outcomes.

Surface-level analysis focusing solely on labor substitution dramatically underestimates transformation value, missing the complex web of benefits that emerge through systematic change. Comprehensive return on investment calculations must account for multiple benefit streams across different time horizons, recognizing that value creation occurs through system-wide optimization rather than isolated efficiency gains. Direct operational benefits provide immediate returns through labor productivity gains of 25-35\% for routine residential properties, cycle time reduction improving customer satisfaction and enabling volume flexibility, revision rate decreases through automated compliance checking, and elimination of calculation and transcription errors.

Hidden value streams emerge through operational transformation as organizations discover previously impossible capabilities. Supervisory efficiency improves through exception-based management that focuses human attention on high-value decisions. Expert knowledge becomes preserved and democratized as heuristics are encoded in systems accessible to all users. Capacity flexibility enables handling volume spikes without proportional staffing increases, while quality consistency reduces variance across appraisers and geographies. These indirect benefits often exceed direct cost savings but prove harder to quantify in traditional ROI calculations.

Risk mitigation value frequently exceeds operational savings, though its probabilistic nature complicates measurement. GSE representation and warranty relief can be worth 5-10 basis points on conforming loans, translating to substantial value for large originators. Reduced buyback risk from consistent, well-documented valuations prevents costly loan repurchases. Fair lending compliance helps avoid violations carrying both financial penalties and reputational damage. Fraud prevention through pattern detection and anomaly identification protects against losses that can devastate profitability.

Strategic analytics capabilities create competitive advantages that compound over time. Granular portfolio risk assessment enables optimized capital allocation, while market intelligence from aggregated valuation data informs expansion decisions. Product development benefits from detailed property characteristic analysis, and regulatory capital optimization becomes possible through better risk measurement. These strategic benefits create sustainable competitive advantages that justify transformation investments even when operational savings alone might not.

Technology-driven unbundling fundamentally reshapes market structure as the integrated appraisal model fragments into specialized components with different economic characteristics. Commoditized data collection faces severe price pressure as barriers to entry collapse, trending toward gig economy models with wages approaching minimum viable levels. Analytical services command premium pricing as professionals demonstrating superior capabilities capture increasing value through volume leverage. Technology platforms exhibit winner-take-all dynamics through high fixed costs and network effects \citep{rochetPlatformCompetitionTwoSided2003}, while boutique specialists find protected niches through differentiation and relationship capital.

First-mover advantages compound through multiple mechanisms creating competitive moats difficult for later entrants to overcome. Data network effects improve model quality with scale, as each processed appraisal enhances algorithms and creates quality gaps competitors struggle to close. Talent concentration accelerates as skilled professionals gravitate toward technology leaders, creating virtuous cycles of improvement. Ecosystem lock-in emerges as workflows integrate deeply, with switching costs compounding through retraining requirements and process disruption. Early movers risk regulatory capture by influencing standards in ways that create subtle but persistent advantages.

The analysis of market structure evolution under AI transformation reveals critical questions requiring ongoing research. Will winner-take-all dynamics dominate or will specialized niches persist? How do network effects interact with regulatory requirements to shape competitive landscapes? What governance mechanisms can prevent harmful concentration while preserving innovation incentives? These questions demand continuous monitoring as markets evolve.

\subsection{Labor Market Implications}

The labor market implications of AI-augmented valuation extend beyond individual job displacement to fundamental restructuring of career paths, skill requirements, and professional identity. These changes exemplify broader patterns identified in the automation and employment literature while exhibiting profession-specific nuances that shape individual and collective responses.

The appraisal profession exhibits classic job polarization patterns identified by \cite{autor2003}, with middle-skill routine jobs facing automation pressure while both high-skill analytical and low-skill manual tasks remain human-centric. This creates an hourglass-shaped occupation structure that fundamentally alters career trajectories and economic prospects. The high-skill segment comprising valuation consultants, review appraisers, and expert witnesses commands premium compensation for complex analytical work requiring deep expertise, sophisticated judgment, and communication skills. Career paths in this segment emphasize specialization and reputation building, with compensation potentially exceeding current averages as volume leverage multiplies earning potential.

Conversely, the low-skill segment of field data collectors and photography assistants performs standardized capture tasks requiring basic technology skills but limited valuation expertise. Compensation faces downward pressure from expanded labor supply and gig economy dynamics, with career advancement requiring transition to analytical roles through education and experience. The hollowing middle of traditional residential appraisers performing full-service valuations faces the greatest pressure, neither specialized enough for premium segments nor willing to accept commoditized wages. This group must adapt or exit, facing not just economic pressure but the psychological challenge of professional identity loss.

Geographic disparities intensify through technology as AI enables separation of locally-bound data collection from location-independent analysis. This creates opportunities for regional arbitrage where skilled analysts in low-cost areas review properties in high-cost markets, though this risks degrading local knowledge essential for quality valuations. Urban concentration of complex properties and specialized services may deplete expertise in smaller markets, while digital divides create valuation deserts in areas with poor connectivity or technology adoption, potentially requiring regulatory intervention.

The aging appraiser demographic complicates adaptation significantly. With average ages exceeding 50, many practitioners face difficult choices between investing in new skills versus pursuing early retirement. Traditional apprenticeship models break down when routine tasks that taught fundamentals face automation, requiring fundamental reimagining of professional education and development. Successful transition requires comprehensive support including educational transformation shifting from measurement techniques to analytical methods, transition assistance through retraining programs and income support, and creation of alternative pathways that preserve valuable expertise while acknowledging changed economics.

Critical questions emerge about how professional cultures adapt to AI augmentation and what interventions most effectively support beneficial transitions. Research from other industries suggests that successful workforce transformation requires early communication, comprehensive support programs, multiple adaptation pathways, and gradual implementation timelines. Organizations must balance efficiency gains with social responsibility, recognizing that abrupt displacement creates costs extending beyond individual hardship to societal disruption.

The temporal modeling challenges for dynamic labor markets add another layer of complexity. Real estate markets exhibit seasonal patterns, economic cycles, and structural breaks that affect employment needs. Research into how AI systems can learn transferable patterns across markets while adapting quickly to regime changes becomes essential for workforce planning. The ability to distinguish signal from noise in volatile periods while maintaining stability determines whether technology creates sustainable employment or destructive churn.

\subsection{Emerging Challenges and Evolution Pathways}

As implementation accelerates, critical challenges emerge that will shape the trajectory of AI-augmented valuation and determine whether transformation enhances or undermines market functioning. These challenges span technical, organizational, and societal dimensions, requiring coordinated responses from multiple stakeholders.

End-to-end uncertainty quantification remains a fundamental technical challenge with profound practical implications. Current methods handle uncertainty at individual stages---measurement accuracy, model confidence, market volatility---but struggle with propagation through complex analytical pipelines. Without comprehensive uncertainty modeling, systems risk false precision that misleads users and regulators. Key research needs include modeling dependencies between uncertainty sources, developing computationally efficient propagation methods, creating interpretable representations for diverse stakeholders, and handling non-stationary uncertainty as markets evolve. Progress in these areas will determine whether AI systems can provide the nuanced risk assessment that sophisticated financial decisions require.

The fundamental limitation of correlation versus causation in AI systems proves particularly problematic for policy-relevant applications. While current models excel at pattern identification, they struggle with causal understanding critical for valuation analysis and market intervention \citep{pearlBookWhyNew2020}. A system might accurately predict that properties near new transit stations appreciate, but without causal understanding, it cannot distinguish whether proximity causes appreciation or whether transit planners strategically locate stations in appreciating areas. Advances require learning causal structures from observational data, handling confounders in non-experimental settings, enabling counterfactual reasoning for scenario analysis, and maintaining causal validity under distribution shifts that characterize evolving markets.

Systemic risks from widespread AI adoption demand proactive management to prevent market instability. Model homogenization represents a primary concern---if market participants converge on similar models trained on overlapping data, correlated errors could amplify rather than cancel. This herding behavior could manifest as synchronized overvaluation during booms or undervaluation during busts, potentially exceeding the systemic risks that human appraisers with diverse perspectives naturally mitigate. Addressing these risks requires diversity requirements mandating heterogeneous models, regular stress testing under extreme scenarios, circuit breakers limiting automated markdown severity, and transparency enabling market monitoring.

The speed of AI-enabled adjustment fundamentally changes market dynamics in ways requiring careful study. While human appraisers update mental models gradually, AI systems can retrain rapidly on new data. This responsiveness improves accuracy during normal conditions but might amplify volatility during transitions. Research into optimal update frequencies, dampening mechanisms, and human oversight triggers becomes essential for maintaining market stability while capturing efficiency benefits.

Adversarial robustness presents novel vulnerabilities as AI systems create new attack vectors for market manipulation. Research on adversarial examples demonstrates how small input perturbations can cause large output changes in neural networks \citep{goodfellow2015, szegedy2014}. For valuation systems, strategic actors might inflate comparable sales through coordinated transactions, flood databases with misleading listings, or exploit known model biases. Developing game-theoretic defenses against such manipulation while maintaining system usability represents an ongoing challenge requiring collaboration between security researchers and domain experts.

The evolution of human-AI collaboration patterns in professional contexts remains poorly understood despite its critical importance. As systems become more capable, the risk of automation bias---over-reliance on AI recommendations---increases. Conversely, excessive skepticism wastes efficiency gains. Research into optimal trust calibration, interface design for effective oversight, and organizational structures supporting collaboration will determine whether AI augments or replaces professional judgment. The development of ``mechanistic interpretability''---understanding not just what models predict but how they reason---offers promising directions for building appropriate professional trust.

International differences in regulatory approaches, market structures, and cultural attitudes toward AI create additional complexity for global organizations. While some jurisdictions embrace technological innovation, others maintain restrictive frameworks prioritizing human oversight. These differences affect competitive dynamics, creating arbitrage opportunities but also compliance challenges. Understanding how to develop AI systems that adapt to local requirements while maintaining global consistency represents an ongoing challenge for multinational participants.

As these challenges illustrate, the transformation of property valuation through AI augmentation is not a deterministic process but one shaped by ongoing choices of technologists, practitioners, regulators, and society. The research frontiers identified throughout this analysis---from technical advances in uncertainty quantification to organizational innovations in human-AI collaboration---will determine whether this transformation enhances professional practice and market efficiency or creates new forms of systemic risk and inequality. Progress demands not just technological sophistication but thoughtful integration of human expertise, regulatory frameworks, and market mechanisms, working together to shape beneficial outcomes from inevitable change.

\section{Conclusions and Future Directions}

This paper has examined the convergence of regulatory evolution and technological innovation in real estate valuation, developing a comprehensive framework for understanding the ongoing transformation of this critical financial function. Our analysis yields insights with implications for theory, practice, and policy that extend beyond the immediate domain to broader questions of AI integration in professional services.

\subsection{Theoretical Contributions}

Our three-layer architecture for AI-augmented valuation systems contributes to multiple theoretical literatures. For information systems research, we demonstrate how domain-specific requirements shape technology architecture in ways that generic frameworks cannot capture. The integration of physical sensing, semantic understanding, and cognitive reasoning provides a template for analyzing other professional domains where AI augmentation must preserve human judgment while enhancing efficiency.

For real estate economics, we bridge traditional valuation theory with emerging technological capabilities, showing how the shift from point estimates to probability distributions better reflects fundamental market uncertainties. Our analysis extends transaction cost economics to professional services, demonstrating how standardization through UAD 3.6 enables new organizational forms while creating novel coordination challenges.

For AI governance literature, we contribute frameworks for evaluating professional AI systems that move beyond generic benchmarks to domain-specific assessment incorporating regulatory compliance, fairness, and human collaboration effectiveness. Our integration of multiple AI ethics concerns---explainability, fairness, uncertainty quantification---demonstrates their interdependence in practice rather than treating them as separate requirements.

\subsection{Practical Implications}

For practitioners, our analysis clarifies both the opportunities and challenges of AI transformation. Success requires not merely adopting new tools but fundamentally reconceptualizing professional roles and value creation. Appraisers must evolve from data collectors to analysts, developing new competencies while preserving essential domain expertise. Organizations must navigate the productivity J-curve with sustained commitment, recognizing that temporary disruption precedes lasting improvement.

For technology developers, we identify critical success factors extending beyond algorithmic performance. Trust emerges through transparency, fairness, and uncertainty quantification---not accuracy alone. Systems must support human oversight and continuous learning rather than attempting to replace professional judgment. Integration with existing workflows and regulatory requirements proves as important as technical capabilities in determining adoption success.

For policymakers, our analysis highlights both transformative potential and risks requiring careful navigation. The technology offers genuine opportunity to address longstanding market failures---inter-appraiser variability, systematic biases, capacity constraints---while creating new challenges around market concentration, workforce displacement, and systemic risk. Regulatory frameworks must evolve proactively rather than reactively, engaging with technology development to shape beneficial outcomes.

\subsection{Limitations and Future Research}

Several limitations bound our analysis and point toward future research needs. First, the rapidly evolving nature of both technology and regulation means specific technical details and regulatory requirements will require continuous updating. Our framework provides structure for understanding changes, but particular implementations will evolve.

Second, limited empirical data on AI system performance in production environments constrains validation of theoretical frameworks. As implementations mature, longitudinal studies tracking actual outcomes---accuracy, bias patterns, user satisfaction, market effects---will prove essential for refining understanding.

Third, our focus on the U.S. market, while allowing deep analysis of specific institutional contexts, limits generalizability. Comparative studies across different regulatory regimes and market structures would illuminate which insights transfer versus remain culturally specific.

Future research should address these limitations while exploring emerging questions:

\begin{itemize}
\item
  How do different AI architectures affect market dynamics and systemic risk?
\item
  What organizational forms best support human-AI collaboration in professional services?
\item
  How can regulatory frameworks balance innovation encouragement with consumer protection?
\item
  What interventions most effectively support workforce transitions while preserving valuable expertise?
\end{itemize}

\subsection{The Path Forward}

The transformation of real estate valuation from craft-based practice to data-driven science augmented by AI represents more than technological upgrade---it fundamentally reconceptualizes how society produces and validates property value information. This transformation's success depends not on technology alone but on thoughtful integration preserving professional values while addressing longstanding market failures.

The window for shaping this transformation remains open but narrows rapidly. Market participants must move decisively while carefully considering broader implications. Early movers will establish patterns---technical standards, business models, professional norms---that prove difficult to change once entrenched. The choices made in the next 2-3 years will reverberate for decades.

Yet urgency must not compromise thoughtfulness. The rush to implement AI should not override careful consideration of impacts on professionals whose expertise remains essential, consumers whose homes represent lifetime investments, and communities where property values shape opportunity. We must resist the technology industry's tendency toward ``move fast and break things'' when those things include careers, wealth, and social equity.

The goal is not replacing human judgment but augmenting it through technology that enhances rather than diminishes professional contributions. In financial markets where property valuations affect credit access, wealth accumulation, and systemic stability, this augmentation must be not merely powerful but provably trustworthy---transparent in operation, fair in application, and robust under uncertainty.

By building on the frameworks developed here, stakeholders can work toward transformation that realizes technology's potential while preserving essential human elements. The future of real estate valuation lies in systems that combine AI's consistency and scale with human expertise in complex judgment, local knowledge, and ethical accountability. This synthesis, rather than replacement, offers the path toward more accurate, fair, and efficient property valuation serving all market participants.

The convergence of UAD 3.6's regulatory catalyst with AI's technological capabilities creates a unique historical moment. How we collectively respond---whether through thoughtful collaboration or fragmented competition---will determine whether this transformation enhances or diminishes the profession's contribution to efficient, fair, and stable real estate markets. The frameworks presented here offer guidance for navigating this transformation, but success ultimately depends on the wisdom, cooperation, and commitment of all stakeholders in building a better future for property valuation.

\section*{Acknowledgments}
The authors acknowledge the use of Claude (Anthropic) and Gemini Deep Research (Google) for literature synthesis, draft efinement, and exploratory analysis. All scientific 
conclusions, critical interpretations, and final text remain the sole responsibility of the human authors, who verified all citations and technical content.

{\footnotesize

}

\clearpage

\section*{Appendix I: 3D Phone Scanning} \label{appendix:3d-scan}

The transformation of property measurement from manual tape measures to sophisticated 3D scanning represents a fundamental shift in how physical spaces are documented for valuation purposes. While the main text examined this evolution through the lens of our three-layer architecture, the practical landscape of available solutions reveals a complex ecosystem of competing technologies, business models, and integration strategies that merit detailed examination (\autoref{tab:scan-3D}).

The proliferation of LiDAR-equipped consumer devices, particularly Apple's iPhone and iPad Pro lines, has democratized access to professional-grade scanning capabilities that were previously the exclusive domain of specialized hardware costing tens of thousands of dollars. This technological democratization has spawned an explosion of competing solutions, each attempting to differentiate through unique combinations of capture modality, processing pipeline, and vertical integration. Understanding this competitive landscape proves essential for organizations navigating build-versus-buy decisions and practitioners selecting appropriate tools for specific use cases.

The market segmentation reflects fundamental trade-offs between accuracy, speed, cost, and ease of use that no single solution optimally balances. Hardware-based solutions from established players like Matterport and iGUIDE continue to dominate high-end professional applications where sub-inch accuracy justifies equipment investments exceeding \$5,000. These systems combine specialized sensors—often integrating multiple capture modalities including structured light, time-of-flight LiDAR, and high-resolution photography—with proprietary processing pipelines refined over millions of scans. Their business models typically involve recurring revenue through cloud processing fees, creating ongoing relationships that fund continuous algorithm improvement.

Mobile-first solutions targeting the broader market have evolved along two distinct trajectories. LiDAR-based applications leverage the depth sensors in premium devices to achieve near-professional accuracy while maintaining the convenience of smartphone capture. Companies like Canvas, Polycam, and Locometric have developed sophisticated on-device processing that provides immediate feedback, addressing the critical limitation of cloud-based solutions where users cannot verify complete coverage until hours after leaving the property. Photogrammetry-based approaches, exemplified by CubiCasa and Hover, sacrifice real-time feedback for broader device compatibility, enabling deployment on any smartphone at the cost of deferred processing and potential re-visits for incomplete captures.

The emergence of hybrid workflows reflects market recognition that different valuation scenarios demand different approaches. Desktop appraisals might rely entirely on third-party imagery and public records, while complex commercial properties require professional-grade scanning equipment. Hybrid appraisals split data collection from analysis, with field technicians capturing standardized imagery that certified appraisers analyze remotely. This unbundling of the traditional appraisal process creates opportunities for specialized providers to address specific workflow components rather than attempting comprehensive solutions.

Integration capabilities have become a critical differentiator as the industry shifts from standalone tools to ecosystem components. The ability to export directly to Xactimate for insurance claims, generate MISMO-compliant XML for UAD 3.6 submissions, or integrate with enterprise valuation platforms often matters more than raw scanning accuracy. Companies like DocuSketch and Plnar have built their entire value proposition around seamless integration with existing workflows, recognizing that frictionless adoption trumps technical superiority in enterprise deployments.

The competitive dynamics reveal interesting patterns about market evolution and technology adoption in regulated industries. Despite the theoretical superiority of newer entrants, established players maintain dominant positions through brand recognition, proven reliability, and deep integration with industry workflows. Matterport's claimed 100+ million square feet scanned creates a data moat that improves their algorithms while providing social proof that risk-averse enterprises value. Meanwhile, newer entrants must find niches—forensic accuracy for Recon-3D, appraisal-specific features for Locometric, or photo-only simplicity for Hosta.ai—rather than competing directly with incumbents.

The \autoref{tab:scan-3D} synthesizes publicly available information about 17 major providers, comparing their technological approaches, target markets, and claimed differentiators. This comparison reveals not just the current competitive landscape but also the evolutionary trajectory of the industry as it transitions from hardware-centric solutions to software-defined capabilities, from professional exclusivity to democratized access, and from isolated tools to integrated platforms.

\begin{table*}[htbp]
\caption{Comparison of different commercial indoor 3D Scanning solutions}
\label{tab:scan-3D}
\centering
\footnotesize
\begin{tabular}[]{@{}p{0.09\textwidth}p{0.12\textwidth}p{0.12\textwidth}p{0.12\textwidth}p{0.15\textwidth}p{0.12\textwidth}p{0.13\textwidth}@{}}\toprule\noalign{}
Company
 & Technology Stack
 & Primary Data Input
 & Target Market(s)
 & Key Features
 & Integration Capabilities
 & Stated Accuracy / Differentiator
 \\
\midrule\noalign{}
Matterport & Proprietary photogrammetry + depth + LiDAR & Pro2, Pro3, BLK360, 360\textdegree{} cams, phone & Real Estate, AEC, Insurance & 3D tours, floor plans, BIM files & Autodesk, Procore, AWS IoT, others & ``Fraction of an inch'' accuracy  \\
Docusketch & AI on 360\textdegree{} camera photos & 360\textdegree{} images + manual input & Restoration, Appraisal, Home Warranty & Floor plans, 360\textdegree{} virtual tours, time-tracked mode & Xactimate  & Fast scan-to-sketch pipeline with 360\textdegree{} camera \\
Cotality & Image Analytics, AI, CoreAI & Property imagery + public records & Lenders, Servicers, Reviewers & 250+ data attributes, fraud detection, image QA & Mercury, CoreLogic, valuation pipelines  & Not a scanning tool; backend analytics stack \\
Xactimate Sketch & Manual/automated sketching & Hand-drawn or imported plans & Adjusters, Restoration & Roof and floor plan sketching with estimate integration & Compatible with Plnar, Docusketch, CoreLogic & Industry standard in claims; input-dependent accuracy \\
Hover & AI + photogrammetry & Exterior/interior phone photos & Roofing, Construction, Remodeling & 3D models, ``to-the-inch'' exterior dims & Verisk, CoreLogic, PDF/XLS/BIM export & ``To-the-inch'' roof + siding takeoffs \\
Polycam & LiDAR + photogrammetry + RoomPlan + drones & iPhone/iPad, drone photos & AEC, Real Estate, Forensics & ``Space'' mode mesh + floor plan, object capture & Blender, Unreal, SketchUp, AutoCAD  & Captures mesh and floor plan in a single scan \\
MagicPlan & AR (RoomPlan) + Bluetooth laser & iOS camera, laser meter & Contractors, Insurance, Remodelers & Live floor plans, estimate lists, checklist forms & Xactimate\textregistered{}, CoreLogic, Zapier & All-in-one estimate + doc workflow \\
Locometric & Brick Mode (LiDAR) + RoomPlan + Touch Mode & iPhone/iPad, Bluetooth laser & Appraisers, Insurers, RE pros & Multi-mode scan, ceiling modeling, GLA calc & Exports to Symbility, DXF, IFC, PDF  & Built specifically for appraisal-grade scans \\
Metaroom & AI + LiDAR + RoomPlan & iPhone/iPad Pro & Architecture, Lighting Design, FM & Multi-room + ceilings + DXF/IFC export & DIALux, Autodesk, DDScad  & Ceiling geometry modeling + multi-floor workaround \\
weScan & LiDAR + cloud photogrammetry & iPhone/iPad Pro & Real Estate, Insurance & 2D/3D floor plans, ANSI/NEN2580 reports & Floorplanner & ANSI-compliant reports \\
Recon-3D & Sensor fusion (LiDAR + photogrammetry) & iOS LiDAR devices & Forensics (crime/crash scenes) & AprilTag scaling, e57 point clouds & Standard forensic tools & Evidence-grade scale accuracy \\
Cubicasa & AI + smartphone + human QA & Any phone & Appraisers, Agents, Photographers & 2D/3D floor plans, GLA report, home tour & Aryeo, HD Photohub, API & ANSI-compliant, 5-minute scan on any phone \\
iGUIDE & 360\textdegree{} camera + LiDAR & iGUIDE PLANIX device & Real Estate, AEC, Insurance & 3D walkthrough, 3,000 sq ft in 15 min, ANSI & RVT, DWG, ESX, Floorplanner  & Professional-grade kit, sub-1\% measurement error \\
Plnar & AI on phone photos & Phone camera (no LiDAR) & Claims Adjusters, Carriers & 3D models, claim-ready data, annotated views & CoreLogic, Verisk (Xactimate)  & Virtual adjusting solution for insurance \\
Hosta.ai & AI via cloud, no app required & 2--6 room photos via URL & Claims, Renovation, Home Services & Floor plans, materials, condition reports & Embeds via API  & Scan from photos + advanced item recognition \\
Canvas & LiDAR + custom CV stack & iPhone/iPad Pro & Design, Remodeling, Interior Architecture & Scan-to-CAD with layered formats & Revit, Chief Architect, AutoCAD & CAD-ready accuracy; 20+ patents \\
ImageSpace & On-device SfM (Structure-from-Motion) & iPhone/iPad Pro & AEC, Real Estate, Prosumers & PDF/3D export, whole-home scanning, photo search & PDF export, 3D share & SfM minimizes drift on-device---no cloud needed \\
\bottomrule\noalign{}
\end{tabular}
\end{table*}

\subsection*{The Absence of Scientific Validation}

Despite the critical role of measurement accuracy in property valuation, no peer-reviewed scientific studies systematically evaluate the real-world performance of major 3D scanning platforms used in appraisal workflows. This absence of independent validation creates an information vacuum filled entirely by vendor marketing materials and anecdotal user reports.

\subsection*{Documented Performance Gaps}

Despite the critical role of measurement accuracy in property valuation, no peer-reviewed scientific studies systematically evaluate the real-world performance of major 3D scanning platforms used in appraisal workflows. This absence of independent validation creates an information vacuum filled entirely by vendor marketing materials and anecdotal user reports.

\subsection*{Documented Performance Gaps}

Limited independent testing reveals a consistent pattern: \textbf{actual measurement errors routinely exceed marketed tolerances by 5-10x}. The gap between vendor claims and documented field performance includes:

\begin{itemize}
\item \textbf{Matterport}: Markets ``within 1\% accuracy'' $\rightarrow$ Reality: 7-16 inch errors (3-5\% deviation) \citep{insiderealestatephotography2025, wegetaround2024}
\item \textbf{CubiCasa}: Claims ``95-97\% accurate'' $\rightarrow$ Reality: 3-17 inch errors (2-5\% deviation) \citep{insiderealestatephotography2025, appraisersforum2024}
\item \textbf{Canvas}: Advertises ``1-2\% for most measurements'' $\rightarrow$ Reality: Up to 20\% errors reported \citep{chieftalk2024, canvas2024}
\item \textbf{Polycam}: States ``up to 98\% accuracy under optimal conditions'' $\rightarrow$ Reality: Highly variable, degrades significantly with lighting and surface conditions \citep{polycam2024, engineering2024}
\end{itemize}

The discrepancy stems from vendors reporting best-case laboratory results (1$\sigma$ confidence) rather than field performance under typical conditions (2$\sigma$ or higher). Environmental factors—direct sunlight, reflective surfaces, complex geometries—systematically degrade accuracy in ways marketing materials fail to disclose \citep{scene3d2024, laserscanningforum2024}.

\subsection*{Implications for Practice}

\textbf{Key Recommendation}: Practitioners should assume real-world accuracy of 3-5\% rather than the 1-2\% commonly marketed. This recalibration is essential for:
\begin{itemize}
\item Appropriate uncertainty quantification in AI models
\item Risk assessment in lending decisions  
\item Professional liability management
\item Setting client expectations
\end{itemize}

The lack of standardized testing protocols—analogous to ISO standards for industrial measurement or RICS protocols for surveying—allows vendors to make aspirational claims without accountability. Until independent scientific validation emerges, \textit{caveat emptor} must guide technology adoption decisions.

\begin{tcolorbox}[
    colback={rgb,255:red,218;green,224;blue,232},
    colframe={rgb,255:red,180;green,190;blue,200},
    title={\textbf{3D Phone Scans}},
    fonttitle=\bfseries,
    boxrule=1pt,
    arc=4pt,
    outer arc=4pt,
    top=10pt,
    bottom=10pt,
    left=10pt,
    right=10pt
]
\textbf{Never accept vendor accuracy claims at face value.} The absence of peer-reviewed validation studies means marketed specifications reflect optimal conditions rarely achieved in practice. Always conduct independent validation for mission-critical applications.
\end{tcolorbox}

\section*{Appendix II: 3D Representations} \label{appendix:3d-repr}

The fundamental limitations of Gaussian-based reconstruction have catalyzed intensive research into alternative primitive representations designed to address the shortcomings of Gaussian primitives. While 3D Gaussian Splatting achieved remarkable success in rendering speed and visual quality, its representational inadequacies for architectural environments have prompted what can be characterized as a Cambrian explosion of alternative primitives, each targeting specific failure modes of the Gaussian formulation.

The core challenge stems from a fundamental representational mismatch: Gaussian primitives are inherently volumetric with soft, continuous density falloffs—ideal for organic shapes and natural scenes but catastrophically inappropriate for the sharp-edged, planar-dominated world of built environments. This mismatch manifests in three critical failure modes that alternative representations specifically address. First, the "cotton ball effect" where hundreds of overlapping Gaussians struggle to approximate a simple flat wall, creating subtle textural artifacts and computational inefficiency. Second, the volumetric inflation problem where thin structures like table legs and window frames become artificially thickened because volume-occupying Gaussians cannot represent essentially one- or two-dimensional structures. Third, the transparency-depth dilemma where glass surfaces create irreconcilable conflicts between geometric accuracy and appearance modeling.

Recent algorithmic innovations have pursued two divergent strategies (\autoref{tab:representation-3D}): enhancing the Gaussian framework through geometric constraints versus abandoning Gaussians entirely for more appropriate primitives. The constraint-based approaches, exemplified by surface regularization techniques and planar-aware methods, attempt to discipline Gaussian behavior through additional optimization objectives that encourage alignment with underlying geometry. These methods show promise but ultimately fight against the fundamental nature of the representation. More radical approaches replace Gaussians with primitives that naturally align with architectural geometry—convex polytopes for sharp edges, beta distributions for bounded surfaces, or a return to classical triangles augmented with differentiable rendering.

\begin{itemize}
\item
  3D Convex Splatting - Uses convex shapes instead of unbounded Gaussians; superior for hard edges and planar surfaces with fewer primitives required. The method achieves an improvement of up to 0.81 in PSNR and 0.026 in LPIPS compared to 3DGS while maintaining high rendering speeds and reducing the number of required primitives \citep{held2024}.
\item
  Deformable Beta Splatting (DBS) - Employs bounded Beta Kernels with adaptive frequency control; 45\% fewer parameters while achieving better memory efficiency for large surfaces. For the first time, splatting-based methods outperform state-of-the-art Neural Radiance Fields, highlighting the superior performance and efficiency of DBS for real-time radiance field rendering \citep{liu2025}.
\item
  Student Splatting and Scooping (SSS) - Uses Student's t distributions with positive/negative densities; achieves comparable results while reducing component count by up to 82\%. The method introduces negative components in an unnormalized mixture model, extending learning into negative density space for enhanced geometric representation \citep{zhu2025}.
\item
  3D Half-Gaussian Splatting - Addresses shape and color discontinuities through truncated kernels; better handling of sharp boundaries. This approach provides a plug-and-play solution that enhances existing 3D-GS methods, achieving state-of-the-art rendering quality without compromising speed \citep{li3DHGS3DHalfGaussian2025}.
\item
  Triangle Splatting - Returns to triangle-based representations with differentiable rendering; achieves over 2,400 FPS and outperforms concurrent methods on indoor scenes. The method is particularly effective in indoor or structured scenes with well-defined surfaces, where triangles can closely approximate geometry and achieve state-of-the-art performance \citep{held2025}.
\item
  Transparent Surface Gaussian Splatting (TSGS) - Specialized framework separating geometry learning from appearance refinement; specifically addresses the transparency-depth dilemma for windows and glass surfaces. TSGS demonstrates state-of-the-art performance with significant improvements in geometric accuracy while improving visual quality \citep{li2025}.
\item
  Spec-Gaussian - Handles anisotropic view-dependent appearance through specialized specular modeling; essential for modern interiors with extensive reflective surfaces. The method addresses the limited ability of spherical harmonics to represent high-frequency information by utilizing anisotropic BRDF modeling for enhanced view-dependent effects \citep{yang2024}.
\end{itemize}

The proliferation of alternatives reveals a deeper insight: no single primitive type can optimally handle all aspects of complex indoor environments. This realization drives emerging research toward hybrid representations that adaptively select primitives based on local geometry—planar primitives for walls, specialized handlers for glass, thin structure representations for architectural details, and traditional Gaussians for organic clutter. The future likely lies not in finding the perfect universal primitive but in orchestrating multiple specialized representations within unified frameworks that leverage each primitive's strengths while mitigating their weaknesses.

\begin{table*}[htbp]
\caption{Comparison of Alternative 3D Representation Methods for Indoor Scene Reconstruction}
\label{tab:representation-3D}
\centering
\footnotesize
\begin{tabular}[]{@{}p{0.10\textwidth}p{0.10\textwidth}p{0.08\textwidth}p{0.08\textwidth}p{0.10\textwidth}p{0.28\textwidth}p{0.08\textwidth}@{}}\toprule\noalign{}
Method
 & Primitives
 & Suitability for Indoor Scenes
 & Geometric Accuracy
 & Photorealistic Fidelity
 & Key Features
 & Reference
 \\
\midrule\noalign{}
PlanarSplatting & Planar Primitives & Very High & Very High & High & Explicitly models scenes with planar surfaces, which are ubiquitous in indoor environments. This leads to extremely fast and geometrically accurate reconstructions of rooms and buildings, with sharp surfaces. & \citep{tan2024} \\
3D Convex Splatting (3DCS) & 3D Convex Polytopes & High & High & Very High & Replaces Gaussians with convex primitives to better represent sharp edges and flat surfaces. This leads to improved geometric fidelity and can achieve high quality with fewer primitives than Gaussian Splatting. & \citep{held2024} \\
Deformable Beta Splatting (DBS) & Deformable Beta Kernels & High & High & Very High & Employs Beta distributions which are more flexible than Gaussians. The deformable nature of the kernels allows for better representation of both geometry and color with fewer parameters, leading to faster rendering. & \citep{liu2025} \\
3D Student Splatting and Scooping (SSS) & Student's t-distributions & High & High & Very High & Uses the more expressive Student's t-distribution and introduces ``scooping'' (negative densities) to carve out detail. This improves parameter efficiency and reconstruction quality. & \citep{zhu2025} \\
Triangle Splatting & Triangles & Very High & Very High & Very High & Merges the advantages of traditional triangle meshes with modern differentiable rendering. It achieves excellent geometric accuracy and is highly compatible with standard graphics pipelines, enabling extremely fast rendering. & \citep{held2025} \\
TSGS & Anisotropic Spherical Gaussians & High & High & High & Specifically designed to reconstruct challenging transparent surfaces by separating geometry learning from appearance refinement. It shows significant improvements in geometric accuracy for objects like glass. & \citep{li2025} \\
\bottomrule\noalign{}
\end{tabular}
\end{table*}

\end{document}